\documentclass[acmsmall]{acmart}

\AtBeginDocument{
  \providecommand\BibTeX{{
    Bib\TeX}}}

\usepackage{hyperref}
\usepackage{url}
\usepackage{amsmath,amsfonts}
\usepackage{listings}
\usepackage{tabularx}
\usepackage{soul}
\usepackage{rotating}
\usepackage{siunitx}
\usepackage{svg}
\usepackage{graphicx}
\usepackage{textcomp}
\usepackage{xcolor}
\usepackage{xspace}
\usepackage{dirtytalk} 
\usepackage{enumitem} 
\usepackage{lineno}
\usepackage{algorithm}
\usepackage{algpseudocode}
\usepackage{verbatim}
\usepackage{caption}
\usepackage{multirow}
\usepackage{color}
\usepackage{tabularray}
\usepackage[framemethod=tikz]{mdframed}
\usepackage[many]{tcolorbox}
\usepackage{array}
\usepackage{hhline} 
\usepackage{tikz}
\usepackage{ifthen}

\definecolor{propmptBoxColor}{RGB}{0,100,164}

\mdfdefinestyle{promptBoxStyle}{
linecolor=propmptBoxColor,outerlinewidth=1pt,
frametitlerule=true,frametitlefont=\sffamily\bfseries\color{white},
frametitlerulewidth=1pt,frametitlerulecolor=propmptBoxColor,
frametitlebackgroundcolor=propmptBoxColor,
backgroundcolor=propmptBoxColor!05,
innertopmargin=\topskip,
roundcorner=2pt,innerleftmargin=4pt,innerrightmargin=4pt,innertopmargin=4pt
}

\mdtheorem[style=promptBoxStyle]{prompt}{Prompt}

\newmdenv[
  backgroundcolor=yellow!30,   
  linecolor=yellow,            
  linewidth=2pt,              
  roundcorner=2pt,           
  innerleftmargin=12pt,      
  innerrightmargin=12pt,
  innertopmargin=10pt,
  innerbottommargin=10pt,
  skipabove=8pt,             
  skipbelow=8pt              
]{sampleBox}

\gdef\Sepline{
  \par\noindent\makebox[\linewidth][l]{
  \hspace*{-\mdflength{innerleftmargin}}
   \tikz\draw[thick,dashed,gray!60] (0,0) --
        (\textwidth+\the\mdflength{innerleftmargin}+\the\mdflength{innerrightmargin},0);
  }\par\nobreak}

\newcommand{\name}{\textsc{LLMigrate}\xspace}

\def\BibTeX{{\rm B\kern-.05em{\sc i\kern-.025em b}\kern-.08em
    T\kern-.1667em\lower.7ex\hbox{E}\kern-.125emX}}

\definecolor{main}{HTML}{535454}
\definecolor{sub}{HTML}{ebebeb}
\newtcolorbox{boxH}{
on line,
    colback = sub, 
    colframe = main, 
    boxrule = 0pt, 
    left=0mm,
    top=0mm,
    bottom=0mm,
    right=0mm,
    leftrule = 4pt
}

\newtcbox{\variableBox}[1][green]{on line,
arc=0pt,outer arc=0pt,colback=#1!10!white,colframe=#1!50!black,
boxsep=0pt,left=1pt,right=1pt,top=2pt,bottom=2pt,
boxrule=0pt,bottomrule=1pt,toprule=1pt}

\newtcbox{\titleBox}[1][gray]{on line,
arc=0pt,outer arc=0pt,colback=#1!10!white,colframe=#1!50!black,
boxsep=0pt,left=1pt,right=1pt,top=2pt,bottom=2pt,
boxrule=0pt,bottomrule=1pt,toprule=1pt}

\newtcolorbox{summery}{colback=gray!5!white,
colframe=gray!75!black, left skip=0pt, left=0pt, top=0pt, bottom=0pt, right=0pt,beforeafter skip=4pt}

\authorsaddresses{}
\setcopyright{acmlicensed}
\acmDOI{10.1145/3728975}
\acmYear{2025}
\acmJournal{PACMSE}
\acmVolume{2}
\acmNumber{ISSTA}
\acmArticle{ISSTA098}
\acmMonth{7}
\received{2024-10-31}
\received[accepted]{2025-03-31}

\begin{document}

\title{Automated Test Transfer across Android Apps using Large Language Models}

\author{Benyamin Beyzaei}
\orcid{0009-0001-4616-9552}
\affiliation{%
  \institution{University of California at Irvine}
  \city{Irvine}
  \country{USA}
}
\email{bbeyzaei@uci.edu}

\author{Saghar Talebipour$^*$}
\orcid{0000-0002-2082-7334}
\affiliation{%
  \institution{University of Southern California}
  \city{Los Angeles}
  \country{USA}
}
\email{talebipo@usc.edu}

\author{Ghazal Rafiei$^*$}
\orcid{0009-0001-5319-3489}
\affiliation{%
  \institution{University of Southern California}
  \city{Los Angeles}
  \country{USA}
}
\email{grafiei@usc.edu}

\author{Nenad Medvidovi\'{c}}
\orcid{0000-0002-1906-4878}
\affiliation{%
  \institution{University of Southern California}
  \city{Los Angeles}
  \country{USA}
}
\email{neno@usc.edu}

\author{Sam Malek}
\orcid{0000-0001-6152-7402}
\affiliation{%
  \institution{University of California at Irvine}
  \city{Irvine}
  \country{USA}
}
\email{malek@uci.edu}

\begin{abstract}
The pervasiveness of mobile apps in everyday life necessitates robust testing strategies to ensure quality and efficiency, especially through end-to-end usage-based tests for mobile apps' user interfaces (UIs). However, manually creating and maintaining such tests can be costly for developers. Since many apps share similar functionalities beneath diverse UIs, previous works have shown the possibility of transferring UI tests across different apps within the same domain, thereby eliminating the need for writing the tests manually. However, these methods have struggled to accommodate real-world variations, often facing limitations in scenarios where source and target apps are not very similar or fail to accurately transfer test oracles. This paper introduces an innovative technique, \name, which leverages Large Language Models (LLMs) to efficiently transfer usage-based UI tests across mobile apps. 
Our experimental evaluation shows \name can achieve a 97.5\% success rate in automated test transfer, reducing the manual effort required to write tests from scratch by 91.1\%. This represents an improvement of 9.1\% in success rate and 38.2\% in effort reduction compared to the best-performing prior technique, setting a new benchmark for automated test transfer.

\end{abstract}






\maketitle
\def\thefootnote{*}\footnotetext{These authors contributed equally to this work}\def\thefootnote{\arabic{footnote}}

\section{Introduction}
\label{Sec_Introduction}
Testing mobile apps’ user interfaces (UIs) is crucial and ensures a seamless user experience across different devices and platforms. 
Manually testing mobile UIs requires significant time and effort and is prone to human error. Many approaches have been proposed in recent years to address these issues and automate the testing process~\cite{li2006effective, kaasila2012testdroid,
su2017guided,mao2016sapienz, amalfitano2012using, gu2019practicalAPE,
dong2020time, wang2020combodroid, hu2018appflow, Monkey, FirebaseRobo, MobiGuitar, FSE12Concolic, OOPLSA13MinimalRestart, PUMA, ICST18QBE, ISSTA13TargetedEvent, MSR15LanguageModelGUI, DynoDroid, EvoDroid, Sapienz, SIG-Droid, FASE13Greybox, ICST14OracleGeneration, ICSE17AndroidWear, Stoat, ICSE17TextInput, ISSTA16MonkeySee}. Many of these automated testing approaches focus on crash detection or maximizing certain criterion such as activity coverage~\cite{mao2016sapienz, wang2020combodroid, su2017guided,
Moran2016CrashScope, gu2019practicalAPE, dong2020time}. However, recent studies~\cite{linares2017developers, Kochar2015Study, haas2021manual} have indicated that developers prefer UI tests that target specific functionalities of an app. For instance, in a shopping app, the tester wants to ensure that a user is able to successfully register, search for a  product, and add the product to the shopping cart.

\textit{Test Transfer}---also known as \textit{Test Migration}
---is a solution proposed in recent years for automatically creating new usage-based tests for a mobile app by using already existing tests from a similar app~\cite{8952228,ASE19Orso}. The concept of test transfer is based on the idea that apps within a specific domain, such as shopping, mail clients, or web browsers, despite potential differences in appearance and the way they are programmed, share very similar functionalities. Therefore, it is possible to use the already existing usage-based tests from one app to automatically create tests that exercise analogous functionalities in another app within the same domain. For instance, use a test for validating the login functionality in \textit{Geek} 
, a shopping app, to automatically create tests for validating the login functionality in other popular shopping apps such as \textit{Yelp} or \textit{Zalando} \cite{zalando}.

The existing test transfer solutions can be divided into two categories: (1)~similarity-based approaches, which focus on matching events between the source and target apps using a distance metric~\cite{behrang2019atm, behrang2018test, lin2019craftdroid, mishra2023image, khalili2024semantic, liu2022test, mariani2021evolutionary, liu2024enhancing}, and (2)~classification approaches that define the test transfer problem as a classification task \cite{hu2018appflow}.

The similarity-based techniques have demonstrated high effectiveness in cases where the source and target apps are very similar.
However, these approaches face significant limitations in more complex cases, especially where a straightforward one-to-one correspondence between the event sequences representing specific functionalities in the source and target apps does not exist~\cite{liu2024enhancing, Zhao_2020}. This limitation is mainly due to the fact that these approaches focus on directly mapping events between the source and target apps. As a result, they encounter limitations in adapting to cases where such direct mappings do not exist, which is often the case in real-world apps. 

The only existing machine learning-based method, AppFlow~\cite{hu2018appflow}, 
has shown to be dependent on the categories of apps that are used for its training, hampering its generalizability. This technique also requires considerable manual effort to adapt to a new app category, further limiting its usefulness.

\looseness=-1
Recent advances in AI have led to the development of Large Language Models (LLMs), which have shown significant promise in automating various software engineering tasks such as code generation~\cite{liu2024your}, code summarization~\cite{ahmed2022few}, and software testing~\cite{wang2024software}.  
These models, trained on a large amount of data, are capable of accurately understanding the semantics of user interfaces, including screens and widgets~\cite{liu2024make}.
Intrigued by these results, we set out to investigate the degree to which the rich semantics embedded in LLMs can be applied to advance the state-of-the-art in test transfer.

This paper proposes \name{},  the first approach to employ multimodal LLMs for transferring UI tests across mobile apps, offering an end-to-end solution without requiring access to the app's source code. 
The proposed technique relies solely on dynamic analysis of the app, as static analysis typically makes the approach less practical for use with proprietary apps for which the source code is not available. 
This approach takes the binaries of two apps---the source and target apps---along with a UI test originally created for the source app as its inputs and generates a corresponding test that targets the same functionality in the target app. 

\name's transfer process consists of two main steps: (1)~\textit{Source Abstraction}, and (2)~\textit{Test Migration}. In the source abstraction phase, the source test is converted to a version represented in natural language, which we call the \textit{Abstract Source Test}. This is achieved by executing the source test on the source app to extract the required features and consulting with the LLM to understand the semantics of the source test. During the second step, test migration, the abstract source test in natural language, which was generated in the first phase, is adapted to the target app. This is achieved by dynamically exploring the target app, extracting features at each step, and selecting the optimal actions with the help of LLM. 

We evaluated \name{} by applying it to transfer UI tests for apps in five different app categories. In each category, tests were transferred across different apps, resulting in a total of 120 transfers. Both the tests and the apps are reused from the dataset introduced by \textsc{CraftDroid}~\cite{lin2019craftdroid}. The transfer outcomes are evaluated based on the precision, recall, reduction, and success rate metrics, and the results were compared to those obtained from the existing techniques evaluated on the same benchmark~\cite{lin2019craftdroid, temdroid, liu2024enhancing}. \name achieved a 97.5\% success rate, reducing the manual effort required to write tests from scratch by 91.1\%. This represents an improvement of 9.1\% in success rate and 38.2\% in effort reduction compared to the best-performing technique previously evaluated on the \textsc{CraftDroid} benchmark. Notably, \name attains these gains with an average transfer time of 247 seconds, surpassing the efficiency of prior methods.
\section{Background and Terminology}
\label{Sec_Background}

\begin{figure*}
\centering
  \includegraphics[width=0.8\textwidth]{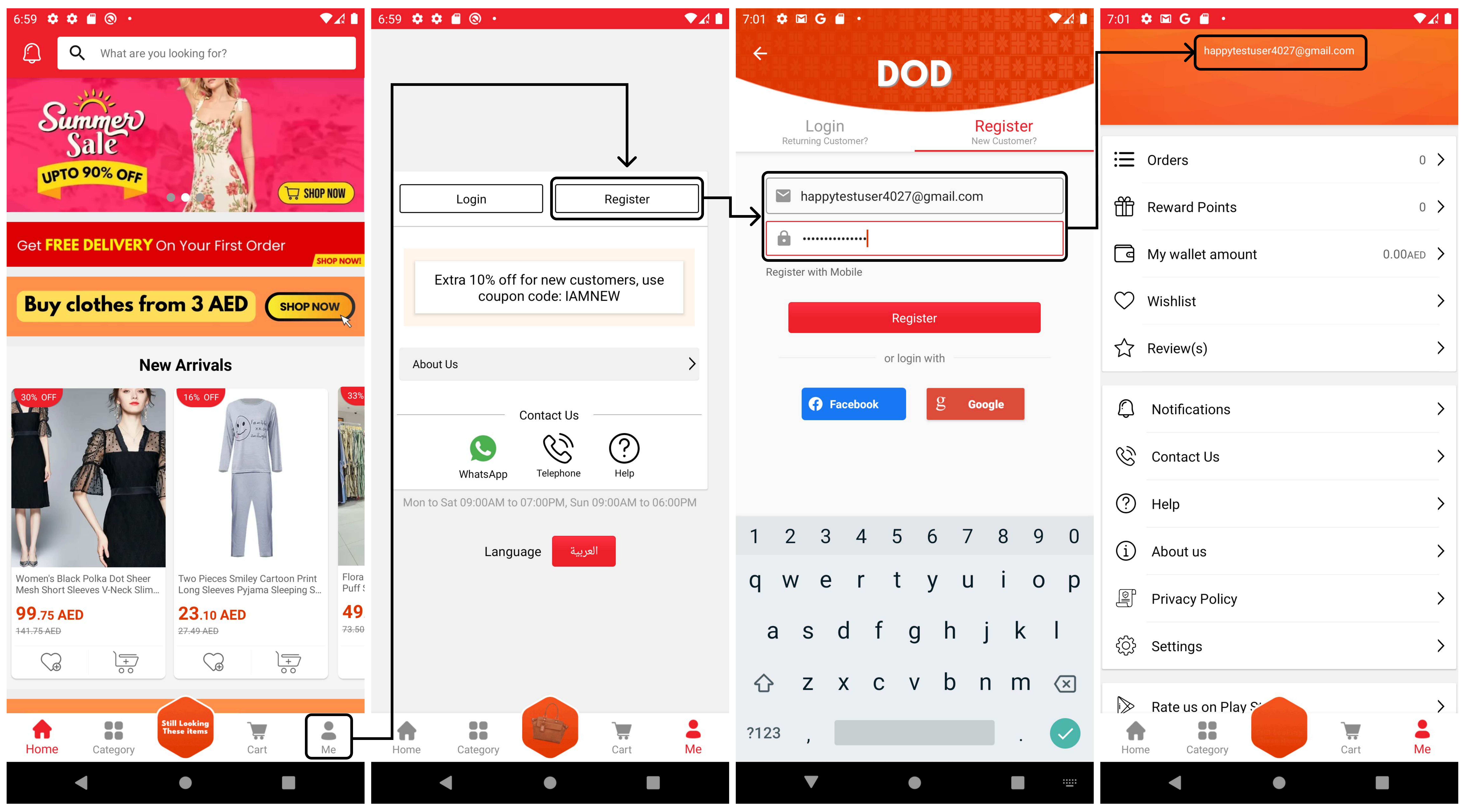}
  \vspace{-5pt}
  \caption{Registration test in DODuae~\cite{dod}.}
  \label{fig:motivation-1}
\end{figure*}

\begin{figure*}
\centering
  \includegraphics[width=\linewidth]{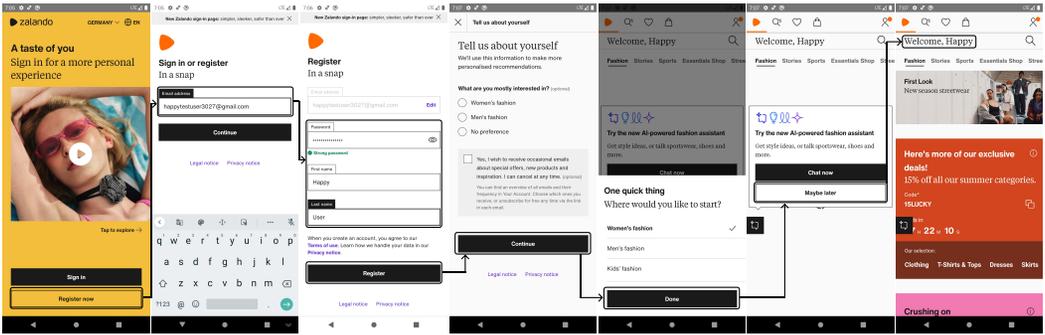}
  \vspace{-15pt}
  \caption{Registration test in Zalando~\cite{zalando}.}
  \label{fig:motivation-2}
\end{figure*}

Figure~\ref{fig:motivation-1} depicts the required steps to register a new user on DODuae, a popular shopping app, while Figure~\ref{fig:motivation-2} demonstrates the same on Zalando, another widely used shopping app. As outlined in Section 1, the core concept of test transfer is to leverage an existing test from one app, such as DODuae, to automatically generate a test that targets the same functionality in another app, in this specific example, Zalando. We use this example to introduce the concepts relevant to our approach.

The \emph{source app} is the app for which an existing UI test is available—in this case, DODuae. This test, called \emph{source test} (e.g., the registration test on DODuae), will be transferred to the \emph{target app}, an app in the same domain as the source app, which in this example is Zalando. The resulting test after the transfer, which is the final outcome of the transfer technique, is termed the \emph{target test} (e.g., the registration test on Zalando). To evaluate the effectiveness of our approach, we compare the target test against a manually created test for the target app, referred to as the \emph{ground truth test}. This ground truth test mirrors the same functionality as the source test and serves as an oracle for measuring the performance of our test transfer technique.

A \emph{UI test} is structured as a sequence of events. The events can be classified into three categories: \emph{UI events}, \emph{system events}, and \emph{oracle events}. UI events represent a user's actions in the app to interact with its user interface, such as taps, swipes, and text inputs. System events are generated by the underlying operating system, such as pressing the back button on Android devices. Oracles, also known as assertions in the relevant literature, are responsible for verifying the expected outcomes at various stages of a test. An oracle in the context of UI testing can be defined as a predicate function $F : (w, c) \mapsto \{\text{\emph{True}}, \text{\emph{False}}\}$, where $w$ represents a widget of the app, and $c$ is a condition to be evaluated pertaining to $w$. The function yields \text{\emph{True}} if the widget $w$ fulfills the criterion $c$; otherwise, the outcome is \text{\emph{False}} resulting in the failure of the assertion.

We represent events in all of these categories as a triple \textit{(action, event\_type, widget)}. Here, the \textit{action} denotes the type of action, such as click, as well as any necessary auxiliary inputs, such as the text input for keyboard events. The \textit{event\_type} can either be \textit{"gui"}, \textit{"oracle"} or \textit{"system"}, specifying the nature of event. The \textit{widget} denotes a target UI element in terms of its attributes. Some events may not be dependent on a specific widget, such as back or scroll, or may not require any auxiliary input.
\section{Approach}
\label{Sec_Approach}

\begin{figure*}[t!]
  \centering
  \includegraphics[width=0.9\linewidth]{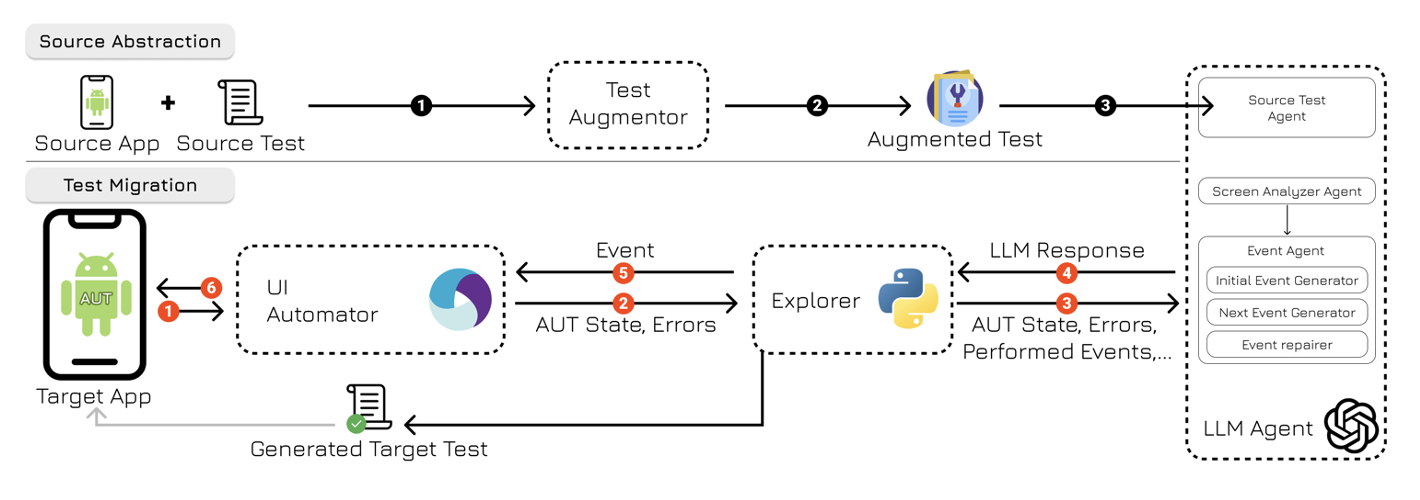}
  \vspace{-10pt}
  \caption{\name's approach overview.}
  \vspace{-10pt}
  \label{fig:approach_overview}
\end{figure*}

Figure~\ref{fig:approach_overview} shows a high-level overview of \name. It takes three primary inputs: 1) a source test, 2) the binary of the source app, and 3) the binary of the target app. Its output is the target test.

\looseness=-1
As shown in Figure~\ref{fig:approach_overview}, \name's approach involves two main phases: \textbf{1)~Source Abstraction}, in which the source test is translated to a natural language representation called the abstract source test, and \textbf{2) Test Migration}, in which the core functionality of transferring the
test to the target app happens by analyzing and understanding the semantics of the target app’s UI. In the following subsections, we will discuss these phases in more detail, and the components involved in each of them.

\subsection{Source Abstraction}
\looseness=-1
During the source abstraction phase, the source test, which is dependent on a specific application, programming language, platform, and testing framework, is translated to an
internal representation called the abstract source test. This internal representation describes the source test in natural language and is independent of any testing framework and programming language. Also, since it is in the form of natural language, the abstract test is easily readable and comprehensible by both humans and LLMs. The translation is
performed in two steps: 1)~\textit{Test Augmentation}, and 2)~\textit{Abstraction}.

\subsubsection{Test Augmentation}\label{sub:augmentation}

The purpose of this step is two-fold. First, it aims to make the test independent of any specific programming language and testing framework. Second, it collects all the useful attributes from the elements that are interacted with in the source test, which might not necessarily be available in the test but can be helpful in understanding its semantics.

In order to achieve this, the source test is executed on a device running the source app event by event. This involves our \emph{Test Augmentor} component communicating with the source app by utilizing the Appium testing framework~\cite{appium}. After the execution of each of the events, this component extracts the UI layout hierarchy of the source app. This layout hierarchy is in the form of an XML tree and represents the UI elements in the current state of the source app. The Test Augmentor then locates the specific widget interacted with during that particular step by analyzing the XML hierarchy, using its locator, such as \texttt{resource-id}. The additional useful textual attributes are then captured from the layout hierarchy for the located widget, such as \texttt{content-desc}, \texttt{text}, or \texttt{class}.

\begin{figure*}[htb]
  \centering
  \includegraphics[width=0.9\linewidth]{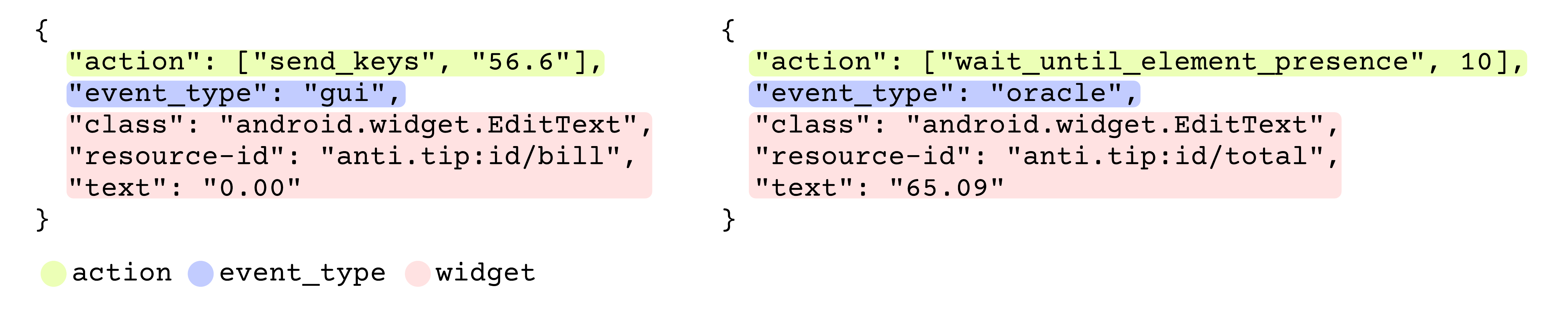}
  \vspace{-5pt}
  \caption{Examples of GUI and oracle events after the test augmentation step.}
  \label{fig:event_sample}
  \vspace{-5pt}
\end{figure*}

Once this information is collected, each event is represented as a triple \textit{(action, event\_type, widget)}, as mentioned in Section~\ref{Sec_Background}. The final augmented test is a sequence of these events, which is independent of any specific programming language or testing framework. Figure~\ref{fig:event_sample} provides examples of GUI and oracle events after the augmentation process.

We support a limited set of UI interactions and primarily focus on two conditions for oracles: (1) the presence of an element, and (2) the invisibility of an element. The set of UI events and oracles supported in our work is consistent with those in the prior test transfer literature~\cite{lin2019craftdroid,temdroid,liu2022test,liu2024enhancing}. Our augmentation module is available publicly on the project repository~\cite{repo}.

\subsubsection{Abstraction}\label{sub:abstraction}
\looseness=-1
In this step, the augmented test, the output of the previous step, is translated to a representation in natural language, called the \emph{Abstract Source Test,}
which serves as a short, one-paragraph summary of the source test that describes each of its UI and oracle events.

To achieve this, we use the \textit{LLM Agent} module, which is responsible for creating prompts and communicating with the off-the-shelf LLMs. To create the prompts for each of the tasks in our approach that require communication with LLM, we have manually created a prompt template that includes the task definition and the necessary structure to introduce each of the input features. This prompt template definition is a one-time manual effort needed for each task, such as test abstraction. Once the template is defined, for every query instance, the \textit{LLM Agent} integrates the variable input features into the prompt template. 

The utilized off-the-shelf LLM accepts two types of prompts: 1)~system prompt, which is a general prompt that familiarizes the agent with the task and provides a general context, and 2)~user prompt, which is task-specific. 
Prompt~\ref{prompt:goal_prompt} represents the user prompt template for the abstraction task, defining the requirements for the abstract test. This includes specifications such as its length and the information it should include.
Note that, in all of the prompts demonstrated throughout the paper, the text presented in color green, represents the variable input features that are substituted in the prompt template for each specific query. For each task, we also provide a system prompt (not shown here) that specifies the context for the LLM. Due to space constraints, the complete prompt details are available in our publicly accessible repository~\cite{repo}.

\vspace{6pt}

\begin{prompt}[Source Abstraction Prompt]
\label{prompt:goal_prompt}
\footnotesize
\noindent This is the test and each step is a JSON object. You should explain what this test does in a single paragraph. Don't add the details of the steps and keep it short but make sure to mention the exact values of inputs and texts. [...] 
\noindent Also explain which step is the final step and which step makes the test complete and emphasize on the functionality that is under test. 
\variableBox{Augmented Test Steps (Generated with \textit{Test Augmentor} module from a test script)}
\end{prompt}
\vspace{2mm}

Once the required prompt is created, the \textit{Source Test Agent} consults the off-the-shelf LLM using its provided API and collects the response. 
The following presents an example of the abstract source test for the registration task of the DODuae app, which was described earlier in Section~\ref{Sec_Background}.

\begin{sampleBox}
\footnotesize
\noindent 
This test contains three oracles and evaluates the registration functionality of the "mobilapp.opencart.doduae" app. It begins by verifying the presence of the dashboard element (oracle), then navigates to the "Me" section (GUI event), and confirms the presence of the register button (oracle). The test proceeds by clicking the "Register" button (GUI event), entering "sample@gmail.com" as the email address and "samplepassword" as the password (GUI events), and clicking the "Register" button again (GUI event). The final step, which completes the test, is an oracle that checks for the presence of the text "sample@gmail.com" to confirm successful registration.

\end{sampleBox}

\subsection{Test Migration}
\label{sec:test-migration}

In this phase, the abstract source test is transferred to the target app by dynamically exploring and analyzing it. It involves the following interconnected components as shown in Figure~\ref{fig:approach_overview}: 1)~\textit{UI Automator}, which is responsible for communicating with the device running the target app to collect information and execute commands on it, 2)~\textit{Explorer component}, which coordinates the main logic of the transfer process, such as tracking the executed events and determining the transfer completion, and 3) \textit{LLM Agent}, which, as discussed in Section~\ref{sub:abstraction}, is responsible for consulting with the LLM. 

 \looseness=-1
 The workflow of the test migration phase is performed as a repetitive loop, executing steps 1 to 6 depicted in orange in Figure~\ref{fig:approach_overview}. 
 In each iteration of the loop, the goal is to find the optimal next event executable in the target app such that the executed sequence in the target app is one step closer to replicating the source test.

The dynamic exploration of the target app always starts from the initial screen that appears when the app is first installed and launched. In the first step, the XML UI layout hierarchy representing the elements in the current screen of the target app is captured by the UI Automator component. Given that this layout hierarchy in its original form contains redundant information, which makes it considerably large, the UI Automator component processes it to clean the layout hierarchy, retaining only the necessary elements identified based on their type. The final set of retained elements includes 15 different widget types, including \texttt{android.widget.EditText} and \texttt{android.widget.ImageButton}, that are widely used across Android apps. This set of widget types is configurable in the tool. 

The complete set can be found in our implementation~\cite{repo}, and if needed, extended. 

We refer to the processed layout hierarchy and a screenshot captured from the app screen as the \textit{Current State} of the app. The current state is the main source of information about the target app at each step of the transfer process. 

Once the app state is collected, it is passed to the Explorer component in the second step. If the Explorer component detects that no previous steps have been executed on the target app, it simply passes the app state to the LLM Agent to identify the optimal initial event to be executed on the current state of the target app in step 3.

\begin{figure*}[t!]
  \centering
  \includegraphics[width=\linewidth]{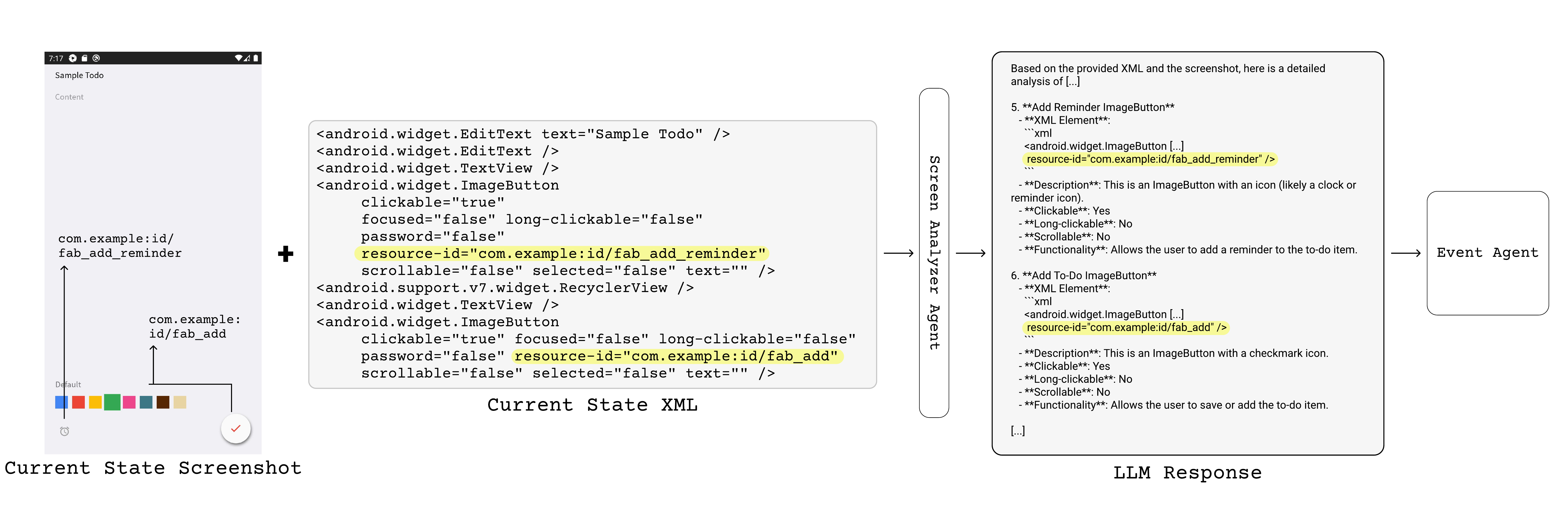}
  \vspace{-15pt}
  \caption{An example of the screen analysis task performed by the screen analyzer sub-agent.}
  \label{fig:screen_analysis}
  \vspace{-5pt}
\end{figure*}

\looseness=-1
In order to detect the optimal event on the target app, the initial step is to understand the semantics of the current app state, including the processed XML UI layout hierarchy and the app screenshot. To achieve this, we have utilized the capabilities of LLMs, as they have demonstrated a strong understanding of structured texts such as XML and HTML~\cite{gur2023real, deng2024mind2web, gur2022understanding, sun2024adaplanner} and are capable of effectively analyzing screenshots of web and mobile apps~\cite{zheng2024gpt}. Specifically, this task is managed by the \emph{Screen Analyzer} sub-agent, as illustrated in Figure~\ref{fig:screen_analysis}. The screen analyzer consults the LLM to analyze both the XML layout hierarchy and the screenshot, matching the widgets within the hierarchy to the captured screenshot. Prompt~\ref{prompt:image_description} represents the specific prompt template used for the screen analysis task. 

The outcome of the screen analysis task is an analysis report, which, as shown in Figure~\ref{fig:screen_analysis}, includes the key elements of the screen along with their locators, natural language functionalities, and the potential actions they can support.

The rationale for using multimodal inputs for the screen understanding task is that relying solely on the textual info obtained from the layout hierarchy can sometimes be insufficient, which may lead to suboptimal event selection. For instance, consider the example of adding an entry in a Todo app, demonstrated in Figure~\ref{fig:screen_analysis}. In this case, two highlighted widgets exist with textual attributes \texttt{fab\_add} and \texttt{fab\_add\_reminder},
respectively. By only relying on textual information, it can be ambiguous to determine which of the two widgets should be used to add a new task, since the term "reminder" is often used interchangeably with "task" or "to-do" in the domain of to-do apps. However, by incorporating the visual information from the captured screenshot, the screen analyzer agent is able to use the images and icons and conclude that the functionality of the second widget is to allow the user to add a reminder to a to-do item.

\vspace{3mm}
\begin{prompt}[Screen Analysis Prompt]
\label{prompt:image_description}
\footnotesize
The attached image is a screenshot of an Android application. Below is the XML representation of the same screenshot: \variableBox{Current State (XML)}

\noindent Your task is to analyze both the image and the XML file. Describe the key widgets you observe in the screenshot and match them to their corresponding XML elements. For each widget, indicate whether:
\par\noindent - It is clickable or long-clickable.
\par\noindent - It is an EditText field, allowing for text input (send\_keys).
\par\noindent - Scrolling is possible within it.
\par\noindent And also explain what is the possible functionality of the widget. \noindent Be detailed in your descriptions and clearly highlight the key widget behaviors.
\end{prompt}
\vspace{3mm}

\vspace{-2mm}
Next, the LLM Agent, specifically the \emph{Initial Event Generator} sub-agent, defines the initial action selection task as a question-answering task and utilizes a specific prompt template for this task. The variable input features for this task include: 1)~The abstract source test, which guides the agent in identifying the next best event; 2)~the XML UI layout hierarchy, which provides information about the existing elements in the target app; and 3) the screen analysis report generated in the previous step. 
Prompt~\ref{prompt:initial_event_prompt} provides the corresponding prompt template for the initial event selection task. Note that in all the prompts presented in the paper, the sections of the prompts written in gray serve as section headings and descriptive summaries. These elements do not exist in the original prompt text but are included in the paper to provide the purpose and content of specific sections of the prompts.

\vspace{2mm}
\begin{prompt}[Initial Event Selection Prompt]
\label{prompt:initial_event_prompt}
\footnotesize
\textbf{\titleBox{Abstract Source Test:}}
This is a sequence of events intended to perform a user interface (UI) test on the source app 
\variableBox{Source App Package Name}
application. We define it as a source test:
\variableBox{Abstract Source Test}
\Sepline
\vspace{1mm}
\noindent\begin{summery}
 \textbf{Task Definition:} This part defines the role of a mobile test engineering assistant responsible for migrating automated tests between Android applications. The focus is on ensuring compatibility with the target app, initializing properly, and adapting test actions based on the target app's current state.
\end{summery}
\noindent You have the initial screen of the target application along with the source test definition above. Now, you are migrating the source test to another application. Given this scenario, which event do you believe is the best to perform next? [...]

\Sepline
 
\vspace{1mm}
\noindent\begin{summery}
\textbf{Event Definition:} This section defines the types of events that can be used when migrating tests: GUI events interact with the app’s interface, Oracle events check the state of the app, and System events handle system-level actions.
\end{summery}

\noindent You can choose between three types of events: 1. GUI events 2. oracle events 3. system events

\noindent 1. GUI events include event\_type as GUI, class of the element, resource-id of the element, action and other necessary fields if they exist like text, hint, content-desc and naf. You should generate this event based on the current state of the target application [...]

\noindent 2. oracle events include event\_type as oracle and the action which can be one of the 

\noindent"wait\_until\_element\_presence" or "wait\_until\_element\_invisible". If you return an oracle event you just need to include event\_type and action fields not anything else. [...]

\noindent 3. system events include event\_type as system [...]
\Sepline 
\noindent \textbf{\titleBox{Application State:}} Here is the initial screen of the current application in XML format: \variableBox{Current State (XML)}

\noindent Here is a description of the current state widgets: \variableBox{Screen Analysis Report (Generated from \textit{Screen Analyzer Agent})}
\Sepline
\noindent \textbf{\titleBox{Possible Actions:}} Instruction for Generating an Action

\noindent Note: Every action should be an array.

\noindent 1. key\_back: Press the system back button. Choose this action when test goal contains back press or back navigation.
example:
["key\_back"]

\noindent [...]

\noindent 9. wait\_until\_element\_invisible: Use this oracle to validate that a particular element is not visible on the screen. The selector type could be "xpath", "content-desc", "id" (resource-id) and "text". you should just include an array inside the action field of the event, alongside with the event\_type. First item inside the array is the action type, second one is the wait time, third is the selector type and the last one is the selector value.
Example:
["wait\_until\_element\_invisible", 10, "text", "a sample text"]: Check if "a sample text" is not visible on the screen for 10 seconds.
\end{prompt}
\vspace{2mm}

The LLM Agent consults the LLM using Prompt~\ref{prompt:initial_event_prompt} to create the initial optimal event, which is expected to be in the JSON format, including the event triple \textit{(action, event\_type, widget)} defined in Section~\ref{Sec_Background}.
Examples for LLM responses that contain an event are shown in Figure~\ref{fig:event_sample}.
In order to mitigate the challenge imposed by the non-determinism of LLMs, we took two specific actions for all the action selection prompts. Firstly, we set the temperature parameter of the LLM to zero to reduce its exploration abilities and make its responses more predictable. Second, to reduce the impact of LLM hallucination, we used a majority algorithm to determine the final generated event. This involves querying the LLM \texttt{n} times repeatedly for each task as a zero-shot prompt, leading to the receipt of \texttt{n} responses in the form of JSON objects. In a JSON object, a key is a unique identifier for a piece of data, and its corresponding value is the data itself. 
To generate the final event, the LLM Agent includes only the keys that appear more than \texttt{m} times among the \texttt{n} received responses and sets the value of these keys to the most frequently occurring value for each included key. Note that \texttt{m} and \texttt{n} are configurable parameters that are set to 2 and 3, respectively, in our current evaluation.

Upon determining the initial event, the LLM Agent sends the generated event to the Explorer component in step 4. The Explorer then records it as an executed event to keep track of it and sends it to the UI Automator component (step 5), so that it can be executed on the target app, enabling it to proceed to the next state (step 6). 

It is important to note that not all the events generated by the LLM Agent are necessarily executable or valid events. Consequently, executing the generated event on the target app may result in one of four different scenarios: \textit{\textbf{1)}}~The UI event is successfully executed on the target app, causing the target device to proceed to a new state, or the condition for the generated oracle event is met on the target app. \textit{\textbf{2)}}~The execution of a UI event or system event leads to an exception, indicating it is not a valid event. This can have reasons such as invalid widget locators created by LLM. \textit{\textbf{3)}}~The condition for the generated oracle is not met in the app, which results in an exception. \textit{\textbf{4)}}~The execution of a UI event will not raise any exceptions, however it will not result in any changes in the target app, indicating it was not a useful step to be included in the target test. Next, we discuss in detail the subsequent steps required to address each of these four scenarios.

\looseness=-1
\textit{\textbf{Scenario 1}} indicates a successful iteration of the migration loop. The migration loop continues to iterate until all the source oracles are transferred to the target app or until a configurable threshold of the number of target events is reached. This threshold is set to three times the number of source test events in our current implementation. Note that, this end condition has the assumption that tests typically conclude with an oracle event, which is often the case in practice. If the end condition is not met, another iteration of the migration loop is initiated. In this scenario, the key difference between the initial and the subsequent iteration is in the event selection prompt. 
For the subsequent steps, the event selection prompt introduces an additional input feature to be integrated into each query alongside the abstract source test, XML layout hierarchy, and screen analysis report, which is the previously executed steps recorded by the Explorer component to guide the LLM Agent in identifying the next optimal event.

\looseness=-1
One of the limitations of the existing test transfer techniques is their failure to handle input value differences between the source and target applications. In real-world scenarios, The required parameters might differ between the source and target apps. For example, some of the required parameters for the source app may not be used in the target app, while some required parameters in the target app may be missing in the source test. For instance, a source test exercising the registration functionality in a shopping app might include a "ZIP code" field not needed in the target app, or the target app might require additional fields like "name" and "family name". Our approach addresses these differences in input requirements by guiding the LLM to generate appropriate values for input fields in case they are not available in the source test while skipping the unnecessary existing input values. The LLM's ability to create appropriate inputs that are not present in the source test based on its understanding of the application's context makes it highly useful for test transfer in contrast to the existing similarity-based techniques, which rely solely on the existing input values in the source test and face shortcomings when additional input values are required.

Additionally, during the event selection process, we incorporate a set of general guidelines, referred to as hints, into the prompts. These hints are designed to help the LLM agent interact more effectively with the application by using a set of specific rules. For instance, one of these hints instructs the agent to ensure that all required fields in a form are filled in before submission. Another hint instructs against selecting repeated actions. Furthermore, the "possible actions" section of Prompt~\ref{prompt:next_event_prompt}, provides additional instructions regarding action types, such as the possibility to interchange \texttt{swipe\_right} and \texttt{long\_click}, across Android applications. We have ensured that all the provided hints are general and contain no app-specific or scenario-specific details.

\looseness=-1
Prompt~\ref{prompt:next_event_prompt} demonstrates the prompt template for 
 the next steps event selection task. Note that the sections that do not contain detailed descriptions are identical to the sections explained in Prompt~\ref{prompt:initial_event_prompt}.

\vspace{2mm}
\begin{prompt}[Next Event Selection Prompt]
\label{prompt:next_event_prompt}
\footnotesize
\textbf{\titleBox{Abstract Source Test}}
\Sepline
\noindent\textbf{\titleBox{Performed Events:}} You have successfully performed these events in the target application so far (performed events array): \variableBox{Performed Events (Captured by \textit{Explorer} module)}
\Sepline
\noindent \textbf{\titleBox{Application State}}
\Sepline
\begin{summery}
\textbf{Task Definition:} This section outlines that during test migration, it is important to prioritize transferring oracle events to verify the correct application state. While exact steps from the abstract source test don't need to be followed, actions should be adapted to the current state of the target app.
\end{summery}
\noindent You are not required to transfer the exact steps in the test goal, just transfer the suitable ones based on the current state.  [...]
\Sepline
\noindent \textbf{\titleBox{Event Definition}}
\vspace{-2mm}
\Sepline
\vspace{1mm}
\noindent\textbf{\titleBox{Values:}} In the target app, you may need to use specific values such as name, email, password, etc. 
Fill out all required fields, but leave optional fields empty.
Use the same values in confirmation fields (e.g., password and confirm password) when required.
If no specific values are provided in the test goal, generate random valid data based on the field names in the target app. Always prioritize using values from the test goal if they match the field in the target app.

\noindent If there are no values to use inside the abstract source test generate random correct values based on the field name in the target app but always prefer using values from the abstract source test for related fields.

\Sepline
\vspace{1mm}
\noindent\begin{summery}
\textbf{Hints:} This part provides general tips for handling events during test migration. Key guidelines include: Base your actions on the "current\_state". Only add oracle events when indicated in the "current\_state". Avoid unnecessary actions like sending keys to already-filled fields. Use attributes like text and class if resource-id is unavailable
\end{summery}
\noindent Some steps and rules that you should follow:

\noindent - If you have already interacted with an element and have not reached the correct oracle, try another action on that element or generate a completely new event. Do not repeat the same events in the already performed events array. [...]

\Sepline
\noindent \titleBox{ \textbf{Possible Actions}}
\end{prompt}
\vspace{2mm}

\textit{\textbf{Scenarios 2}}  and \textit{\textbf{3}} occur when an invalid event is created by LLM, which are easily identifiable by the UI Automator component. This is because executing these events on the target app results in raising an exception. In these cases, again, another iteration of the migration loop begins. However, since the previous iteration was unsuccessful, additional measures need to be taken by Explorer, our primary coordinator, to address the failure. These measures are fourfold. First, the Explorer component removes the previously generated event that resulted in an exception from the list of executed events, as it should not be included in the final transferred test. Second, the Explorer records this event as a dead-end event to ensure that it will not be mistakenly considered a valid event in future iterations. It is important to note that dead-end events are saved based on the current transfer step, as an event might be a dead-end in one step but not in another one. During step 5 of the migration loop, the Explorer always checks to ensure that a suggested event by the LLM Agent has not been previously identified as a dead-end event.
Third, the Explorer backtracks one step in the execution to undo the consequences of the last invalid generated event. The backtrack is achieved by restarting the app, clearing the cache, and re-executing all recorded events except the last invalid one. 
Finally, the Explorer needs to inform the LLM Agent that the generated event in the previous iteration was invalid. This results in the LLM Agent utilizing a specific prompt, asking the off-the-shelf LLM to repair its previous attempt. 

\looseness=-1
Prompt~\ref{prompt:repair_event_prompt} shows the sections of the repair event selection prompt that are different from Prompt~\ref{prompt:next_event_prompt}. This prompt specifically includes a feedback section that consists of the previously generated incorrect event and the exception that was raised as a result of executing that event. The rationale behind including this feedback is based on prior research in program repair and analysis, which has shown that LLMs have a strong ability to understand bugs and exceptions when they are provided with detailed feedback~\cite{bouzenia2024repairagent}. Therefore, to enhance the handling of cases in which an exception is raised, we use a Chain-of-Thoughts (CoT) approach by including the exception description and prior responses within the prompt. This method has been shown to improve the reasoning capabilities of LLMs~\cite{wei2022chain}.

\vspace{2mm}
\begin{prompt}[Repair Event Prompt]
\label{prompt:repair_event_prompt}
\footnotesize
\textbf{\titleBox{Abstract Source Test} + \titleBox{Performed Events} + \titleBox{Application State}}
\Sepline
\noindent \titleBox{\textbf{Feedback:}} You have been asked for generating an event, this is the last generated event and it has already been attempted and failed, throwing an exception. This event should not be recommended again in your response: \variableBox{Last Wrong Event}

\noindent This event is not correct because of this exception: \variableBox{Last Exception}
\Sepline
\noindent \textbf{\titleBox{Task Definition} + \titleBox{Event Definition} + \titleBox{Values}}
\Sepline
\noindent 
\looseness=-1
\textbf{\titleBox{Hints:}} 
- If you have already interacted with an element and have not reached the correct oracle, try another action on that element or generate a completely new event. Do not repeat the same events from the already performed events array.

\noindent - When your last wrong event is a send key action but the exception indicates that you cannot set the element and you are interacting with the wrong element, you should try clicking on that element first before sending keys. Fix the last wrong event by changing the action to click. [...]
\Sepline
\noindent \textbf{\titleBox{Possible Actions}}
\end{prompt}
\vspace{2mm}

\textit{\textbf{Scenario 4}}, where the execution of the generated event does not result in any changes in the target app's state, is not identifiable by the UI Automator as it does not raise an exception. To detect this scenario, at the beginning of each migration loop iteration, the Explorer component compares the target app state received from the UI Automator with its previous state.
If the two states are identical, indicating that the generated event was not useful, the Explorer takes the same steps required for scenarios 2 and 3 mentioned above, including backtracking and utilizing the repair event selection prompt (Prompt~\ref{prompt:repair_event_prompt}) in the iteration. 
\looseness=-1
To prevent the migration process from becoming stuck on a specific screen and continuously generating invalid events, a threshold is established for the number of incorrect events generated in each step of the transfer process. 
Upon reaching the threshold of unsuccessful attempts, which is set to three attempts in the current implementation, the Explorer initiates a backtrack. The last performed event is also removed from the executed events in Explorer's record, and it is marked as a dead-end event.

As previously mentioned, the migration loop continues until all the source oracles are transferred to the target app. At this point, the Explorer creates the final output from the recorded executed events, resulting in the generated target test. Note that the generated target test is presented as an augmented test, which is in the form of a triple \textit{(action, event\_type, widget)}, as discussed in Section 2. In this representation, the widget attribute may contain different selectors, such as \texttt{resource\_id} or \texttt{text}, any of which can be used for widget interaction. As previously mentioned, the UI Automator component is responsible for executing the events of the transferred test on the device which includes prioritizing the widget selectors to be utilized for the event execution. 
\section{Evaluation}
\label{Sec_Evaluation}
In this section, we detail the evaluation of \name, focusing on how effective it is in transferring UI tests across mobile apps. Our evaluation aims to answer the following four research questions:

\begin{enumerate}[leftmargin=1.1cm,label=\bfseries RQ\arabic*.]
    \item How effective is \name in accurately transferring UI tests across real-world mobile apps?
    \item How useful are the tests transferred using \name?
    \item What are the time and cost implications of using \name for transferring tests?
    \item How practical is \name for transferring tests on today’s popular apps?
    \item How well does \name perform in transferring tests across apps that are not previously seen by LLMs?
\end{enumerate}

\subsection{Experimental Setup}\label{llm_experimental_setup}

\looseness=-1
\name is designed to be platform-agnostic and, therefore, capable of transferring tests across various devices and platforms. Our current implementation focuses on transferring tests across Android apps. For our evaluation, we utilized the publicly available dataset introduced by the authors of \textsc{CraftDroid}~\cite{lin2019craftdroid}. As shown in Table~\ref{table_subjects_2}, this dataset contains tests from five different app categories: Browser, To Do List, Shopping, Mail Client, and Tip Calculator. The advantages of utilizing this dataset are two-fold. First, it enables the evaluation of our developed approach within the context of real-world apps. Second, since this dataset has also been utilized as the benchmark for evaluating the other existing test transfer techniques, such as \textsc{CraftDroid}~\cite{lin2019craftdroid}, \textsc{TEMdroid}~\cite{temdroid}, \textsc{TRASM}~\cite{liu2022test}, and \textsc{TREADROID}~\cite{liu2024enhancing}, it enables us to effectively compare our technique against the state-of-the-art approaches across various dimensions such as accuracy, usefulness, and performance.

It is important to note that we were unable to utilize the \textsc{CraftDroid} dataset in its original
form because, for some of the apps, the versions that were used in this dataset are no longer functional. In these cases, we used an updated version of the apps, provided the following two constraints hold: 1) a functional and supported version of the app is available, and 2) the update does not alter the test flow from the original version utilized in the \textsc{CraftDroid} dataset, such as requiring additional steps like CAPTCHA. In the cases where these conditions were not met, we removed the non-functional app from the dataset. This resulted in reusing 19 out of the 25 apps from the original \textsc{CraftDroid} dataset. Table~\ref{table_subjects_2} demonstrates the final set of the subject apps as well as the versions used in our evaluation.

\begin{table}[ht]
\centering
\caption{Subject apps.}
\vspace{-5pt}
\label{table_subjects_2}
\resizebox{\textwidth}{!}{
\begin{tabular}{cccccc}
	\hline
	\textbf{Category} & \textbf{c1-Browser} & \textbf{c2-To Do List} & \textbf{c3-Shopping} & \textbf{c4-Mail Client} & \textbf{c5-Tip Calculator} \\ \hline
	&
        Lightning (5.1)&
        Minimal (1.2) &
        Geek (2.3.7) &
        K-9 Mail (6.603) &
        Tip Calculator (1.1) \\

        \textbf{Apps}&
	Browser for Android (6.0)&
        Clear List (1.5.6) &
        Yelp (10.21.1) &
        Mail.Ru (14.117.0)&
        Tip Calc (1.11) \\
        
        \textbf{(Versions)}&
	Privacy Browser (2.1) &
        To-Do List (2.1) &
        &
        myMail (14.97.0)&
        Simple Tip Calculator (1.2)\\

        &
	FOSS Browser (5.8)&
        Shopping List (0.10.1)&
        &
        &
        Tip Calculator Plus (2.0)\\

        &
	Firefox Focus (6.0)& 
        &
        &
        &
        Free Tip Calculator (1.0.0.9)
        \\ \hline
	
	\end{tabular}
}

\end{table}

In the resulting dataset, there are tests for validating at least two of the main functionalities provided by apps in each category, as shown in Table~\ref{table_tests}. In the presented table, categories are represented by \emph{c}, and functionalities under tests are represented by \emph{t}. For example, \emph{c1/t1} represents the first tested functionality for the first app category, Browser, which is \emph{Access website by URL}. We conducted evaluations on a total of $120$ transfers and manually evaluated the transfer results.

\begin{table}[t]
\centering
\caption{Test cases for the proposed functionalities.}
\label{table_tests}
\vspace{-5pt}
\resizebox{0.7\textwidth}{!}{
{\renewcommand{\arraystretch}{1.0}

\begin{tabular}{l c c c }
	\hline
	\multicolumn{1}{c}{\multirow{2}{*}{\textbf{Functionality}}} & \vtop{\hbox{\strut \textbf{\#Test}}\hbox{\strut \textbf{Cases}}} & \vtop{\hbox{\strut \textbf{Average\#}}\hbox{\strut \textbf{Total Events}}} & \vtop{\hbox{\strut \textbf{Average\#}}\hbox{\strut \textbf{Oracle Events}}} \\ \hline
	c1/t1-Access website by URL & 5 & 3.6 & 1.0 \\ 
	c1/t2-Website navigation involving back button & 5 & 6.6 & 3.0 \\ 
	c2/t1-Add task & 4 & 4.25 & 1.0 \\ 
	c2/t2-Add then remove task & 4 & 6.75 & 2.0 \\ 
	c3/t1-Registration & 2 & 14.5 & 5.0 \\ 
	c3/t2-Login with valid credentials & 2 & 7.0 & 3.0 \\ 
	c4/t1-Search email by keywords & 3 & 5.0 & 3.0 \\ 
	c4/t2-Send email with valid data & 3 & 9.3 & 3.0 \\ 
	c5/t1-Calculate total bill with tip & 5 & 3.8 & 1.0 \\ 
        c5/t2-Split bill & 5 & 4.8 & 1.0 \\ \hline
	\multicolumn{1}{c}{Total} & 38 & 5.9 & 2.0 \\ \hline
	\end{tabular}
}
}
% \vspace{-5mm}
\end{table} 

Our experiments were conducted using a Nexus 5X emulator running Android 6.0 (API 23), aligning with \textsc{CraftDroid}'s evaluation for apps where the original versions were functional. For the updated apps that are incompatible with this older version, we employed Nexus 6a emulators running Android 10.0 (API 29).
For our evaluation, we used GPT-4o as an off-the-shelf LLM provided by OpenAI, which can perform reasoning across both visual and textual inputs. We selected this model due to its reasonable cost and superior performance on reasoning and coding benchmarks \cite{GPT-4o}. All tests were conducted through OpenAI’s API, and to accurately measure the transfer time, we tested across various internet connections and VPNs. The transfers were executed on a Mac machine with an 8-core CPU, 10-core GPU, and 16GB of unified memory.

As discussed in Section~\ref{Sec_Approach}, our approach has three adjustable parameters: 1)~maximum wrong tries at the Same Step, 2)~total number of runs for majority voting (n), and 3)~threshold for majority voting to include a field (m). We empirically observed the best-performing values for all these parameters and set them to 3, 3, and 2, respectively, in our evaluation.

\subsection{RQ1. Efficacy of \name{}}

\looseness=-1
For evaluating the efficacy of \name, we utilize the precision and recall metrics introduced and utilized by the existing research targeting test transfer~\cite{lin2019craftdroid, temdroid, liu2024enhancing}. Similar to the definition utilized by the existing research, true positives (TP) are events in the transferred test that exist in the ground truth. False positives (FP) are events in the transferred test that do not exist in the ground truth. Finally, false negatives (FN) are events that exist in the ground truth but are not present in the transferred test.

\definecolor{darkgreen}{rgb}{0.0, 0.5, 0.0}
\definecolor{darkred}{rgb}{0.6, 0.0, 0.0}
\definecolor{darkyellow}{rgb}{0.8, 0.6, 0.1} 
\newcommand{\increased}[1]{
   ( \textcolor{darkgreen}{\tikz[baseline]{
    \draw[darkgreen, semithick, ->] (0,-0.05em) -- (0,0.6em);
} \hspace{0.1em}#1\%})
}
\newcommand{\decreased}[1]{
    ( \textcolor{darkred}{\tikz[baseline]{
    \draw[darkred, semithick, ->] (0,0.6em) -- (0,-0.05em);
    } \hspace{0.1em}#1\%})
}
\newcommand{\same}[1]{
    ( \textcolor{darkyellow}{\tikz[baseline]{
    \draw[darkyellow, semithick] (-0.35em, 0.3em) -- (0.35em, 0.3em);
    } #1\%})
}
\renewcommand{\arraystretch}{1.1}

\begin{table}[b]
\centering
\vspace{-8pt}
\caption{Comparative analysis of precision and recall score metrics achieved by \name, \textsc{CraftDroid}, and \textsc{TREADROID} across different app categories and \textsc{TEMdroid}'s average.}
\resizebox{0.8\textwidth}{!}{
\begin{tabular}{cccccc}
\hline

\multirow{2}{*}{\textbf{App Category}} & \multirow{2}{*}{\textbf{Approach}} & \multicolumn{2}{c}{\textbf{Precision}} & \multicolumn{2}{c}{\textbf{Recall}} \\ \cline{3-6}&& \textbf{GUI Event} & \textbf{Oracle Event} & \textbf{GUI Event} & \textbf{Oracle Event} \\ \hline
\multirow{3}{*}{Browser} & \name & 
    100 \same{}
    & 100 \same{} 
    & 100 \same{}
    & 100 \same{}
    \\
                       & \textsc{CraftDroid} & 90.54 & 100 & 98.52 & 97.50 \\ 
                       & \textsc{TREADROID} & 100 & 100 & 100 & 100 \\
                       \hline
                       
\multirow{3}{*}{To Do List} & \name & 
93.26 \increased{7.9} & 
97.22 \increased{4.9} & 
98.98 \same{} & 
97.22 \increased{5.8} \\ 
                       & \textsc{CraftDroid} & 83.48 & 92.30 & 78.44 & 80\\ 
                       & \textsc{TREADROID} & 85.39 & 91.42 & 98.70 & 91.42 \\
                       \hline
                       
\multirow{3}{*}{Shopping} & \name & 
 100\increased{55.2}& 
 56.25 \increased{27.7} &
  96.87 \increased{32.1} & 
 56.25 \increased{23.4}\\ 
                       & \textsc{CraftDroid} & 44 &28.57 & 64.70 & 35.29\\ 
                       & \textsc{TREADROID} & 44.73 & 27.78 & 51.52 & 32.81\\ 
                       \hline

\multirow{3}{*}{Mail Client} & \name & 
100 \increased{6.8} & 
91.67 \decreased{8.3} & 
100 \same{} & 
91.67 \decreased{8.3} \\ 
                       & \textsc{CraftDroid} & 64.70 & 83.33 & 68.75 & 83.33\\
                       & \textsc{TREADROID} & 93.18 & 100 & 95.34 & 100 \\ 
                       \hline

\multirow{3}{*}{Tip Calculator} & \name & 
100 \increased{5.0} & 
100 \same{} & 
100 \increased{4.0} & 
100 \same{} \\ 
                       & \textsc{CraftDroid} & 77.78 & 75 & 88.15 & 73.17\\ 
                       & \textsc{TREADROID} & 94.94 & 100 & 95.91 & 89.74\\ 
                       \hline
\multirow{4}{*}{Total Average} & \name & 
98.39 \increased{9.6} & 
94.71 \increased{3.1} & 
99.53 \increased{6.0} & 
94.71 \increased{5.7} \\ 
                       & \textsc{CraftDroid} & 78.54 & 83.59 & 85.65 & 82.29\\ 
                       & \textsc{TREADROID} & 88.77 & 91.58 & 93.52 & 87.44\\ 
                       & \textsc{TEMdroid} & 71 & 90 & 93 & 89 \\ \hline
\end{tabular}
}

\label{table_rq1}
\end{table}

Table~\ref{table_rq1} presents a comparative analysis of the precision and recall metrics obtained by \name, \textsc{CraftDroid}~\cite{lin2019craftdroid} and \textsc{TREADROID}~\cite{liu2024enhancing}. Consistent with prior research, we categorized all events into two main types: GUI events and oracle events, with system events classified under GUI events. To ensure a fair and accurate comparison, we aimed to capture results achieved by the other techniques that were derived solely from the 19 subject apps that are currently functional from the \textsc{CraftDroid} dataset, as shown in Table~\ref{table_subjects_2}. This required the detailed analysis of each transferred test for other techniques, similar to the process we used to analyze tests transferred by \name. This was possible for \textsc{CraftDroid} and \textsc{TREADROID}, as the artifacts containing the transferred tests are publicly available for \textsc{CraftDroid}, and we obtained the corresponding artifacts for \textsc{TREADROID} upon communicating with the authors. This was not possible for \textsc{TEMdroid} as the final transferred tests are not publicly available, and we were unable to obtain the research artifacts that would allow us to process it for a fair comparison, even after contacting the authors. Furthermore, we were unable to successfully run \textsc{TEMdroid} as the implementation of certain components was not publicly available. Consequently, for \textsc{TEMdroid}, we used the average metrics reported in their paper. In this case, although the datasets are not identical, the comparison remains informative since our dataset shares 82.6\% (19 out of 23) of the apps. 
  We have not included the metrics obtained by \textsc{TRASM}~\cite{liu2022test} and \textsc{ATM}~\cite{behrang2019atm}, as \textsc{TREADROID} previously benchmarked its obtained results against these techniques and demonstrated superior performance across all metrics~\cite{liu2024enhancing}.

  \looseness=-1
  Note that there may be more than one correct ground truth in transferring a test from a source to a target app. To this end, we manually inspected the transferred tests generated by all the transfer techniques rather than automating the process to ensure a fair comparison between different techniques.

As demonstrated in Table~\ref{table_rq1}, \name was able to achieve a total average precision of 98\% for GUI and 94\% for oracle events. Similarly, \name achieved an average recall of 99\% for GUI events and 94\% for oracle events. These results indicate that \name outperforms all the existing techniques in both average precision and recall metrics for both GUI and oracle events in total. Note that, due to the varying lengths of different scenarios, all reported average metrics are calculated based on the total number of events within each category or across all transfers rather than based on the achieved metric for each individual migration. Furthermore, a detailed analysis of the results for each individual migration is presented in our publicly available repository~\cite{repo}.

For the app categories containing less complicated apps and test flows, such as the browser category, the baseline method demonstrated successful transfer of all events and oracles, and \name was able to achieve similar performance. However, in more complex scenarios, we observe a notable improvement over the baseline. This improvement is largely due to the \name's more advanced comprehension of screen elements, enabling it to surpass the limitations of one-to-one event transfer.

Unlike GUI events for which we observed improvement in both precision and recall across all app categories, we saw degradation in one category in these metrics for oracle events. Upon further analysis of these cases, we realized that in these cases, many oracles are specifically designed to confirm page accuracy prior to executing an action. When LLM can detect a widget on a page and interact with it directly, sometimes it omits the necessary oracle transfers for verification purposes. This may account for a decrease in oracle-related metrics in categories such as mail applications. Earlier methods, which transfer events sequentially, transfer oracles at particular steps, thus benefiting from consistent application flows and yielding marginally higher precision and recall in oracle events. But this is not always the case, and \name design principles allow it to achieve better performance on apps which have different flows.

\vspace{-6pt}
\subsection{RQ2. Usefulness of the Tests Transferred by \name{}}

\looseness=-1
The usefulness of a test is defined by how helpful it is in reducing the manual effort for a human tester. To measure usefulness, we utilized the reduction metric, defined by previous research in this area~\cite{Zhao_2020}. This metric compares the manual effort required to write the ground truth test from scratch to the effort required to manually transform the transferred test to be identical to the ground truth test. 
The manual effort is defined as the Levenshtein distance~\cite{levenshtein1966binary} between the sequence of events of the transferred and ground truth tests. The reduction metric is calculated using the following equation: (\# Ground Truth Events - Manual Effort)~/~(\# Ground Truth Events).

\begin{figure*}[h]
  \centering
  \includegraphics[width=\linewidth]{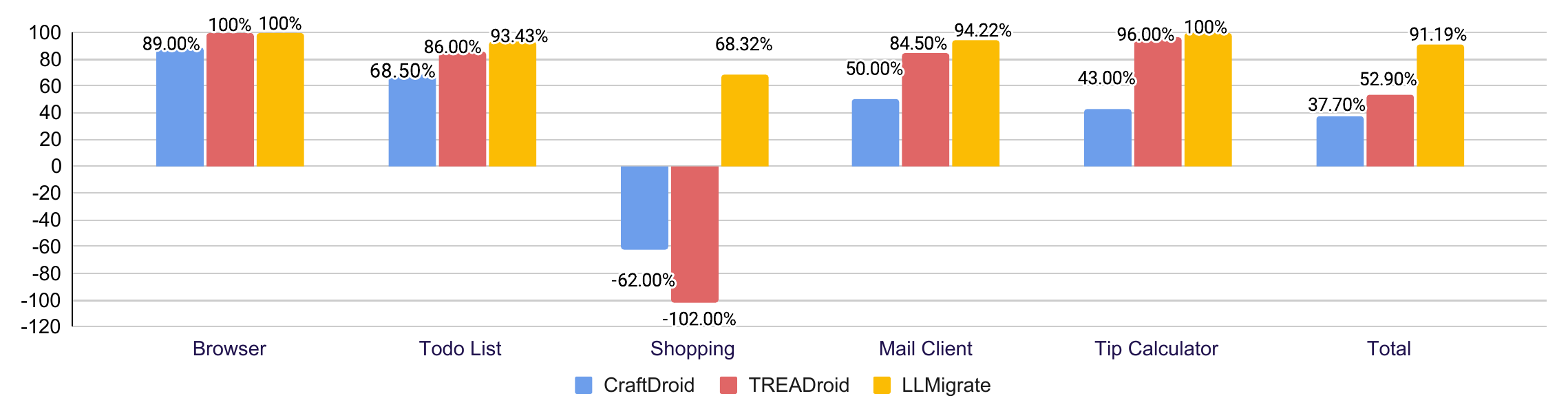}
  \caption{The reduction achieved by \name, \textsc{CraftDroid}, and \textsc{TREADROID} across various app categories.}
  \vspace{-5pt}
  \label{fig:reduction_comparison}
\end{figure*}

Figure~\ref{fig:reduction_comparison} presents the average reduction metric achieved by \name, \textsc{CraftDroid}, and \textsc{TREADROID} across different app categories and in total. Similar to the analysis performed for answering RQ1, we used the subset of the subject apps common across all approaches to ensure a fair comparison and evaluated all the transferred tests manually. Again, we were unable to include the reduction metric achieved by \textsc{TEMdroid} due to the unavailability of their artifacts and not reporting the reduction metric on \textsc{CraftDroid} dataset in the corresponding publication. On average, \name achieved 91\% reduction, demonstrating that \name was able to eliminate more than 91\% of the manual effort required for writing tests, outperforming all the prior techniques by almost 40\% in total average. 

In categories such as shopping, which involve more complex test cases to transfer, previous research ~\cite{lin2019craftdroid, liu2024enhancing} has shown a negative reduction metric, indicating that it can be more efficient to write tests from scratch rather than rely on transfer methods followed by extensive manual edits to the generated tests. This insight underscores the significance of metrics like reduction, which directly reflect the decrease in manual effort, as opposed to focusing solely on precision and recall of the transferred events. Since tools in this domain are fundamentally intended to transfer complete tests to minimize manual work for test engineers, reduction serves as a more relevant measure of a tool’s practical effectiveness.

In our evaluation of \name's usefulness, we utilized another metric called the successful transfer rate. This is a binary metric, assigned a value of 1 (100\%) if the objective of the source test is met in the transferred test and 0 (0\%) if it is not. Detecting if the objective of the test is met is done through manual inspection. 

\begin{figure*}[b]
  \centering
  \includegraphics[width=\linewidth]{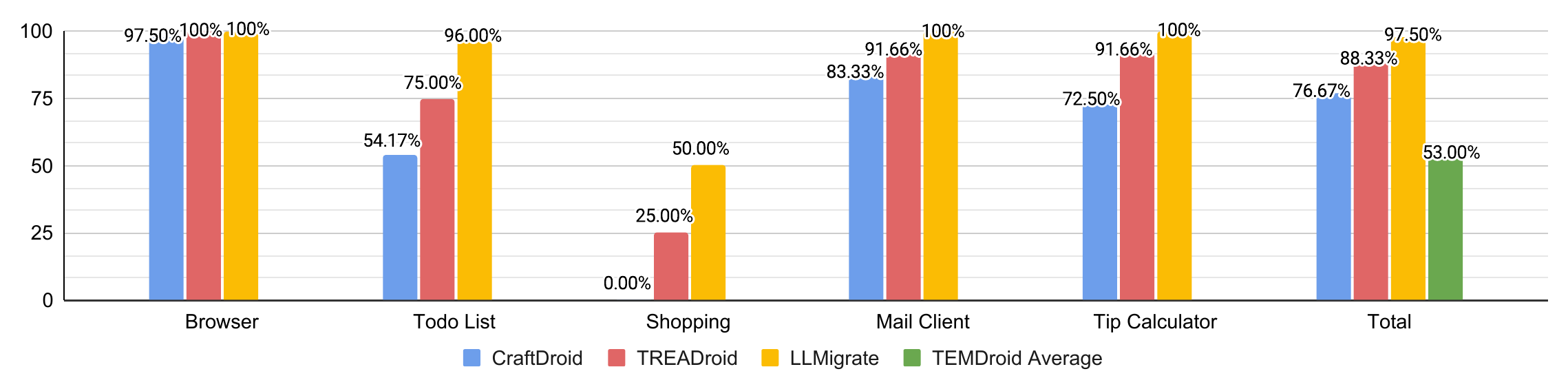}
  \caption{The successful transfer rate achieved by \name, \textsc{CraftDroid}, \textsc{TEMdroid}, and \textsc{TREADROID} across various app categories.}
  \label{fig:success_rate}
\end{figure*}

Figure~\ref{fig:success_rate} presents the successful transfer rate achieved by various techniques across different app categories and in total. On average, \name{} was able to achieve a total of 97.5\% successful transfer rate across all the 120 transfers. This means that, although the transferred test may have additional extra steps not present in the ground truth or reach the objective via a path that was not identical to our specific manually defined ground truth, 97.5\% of the transferred tests successfully met their objectives, including executing the required functionality and asserting the appropriate conditions using the transferred oracles. This shows that \name was able to outperform the existing technique with the highest successful transfer rate, \textsc{TREADROID}, by almost 10\%. Again, for \textsc{TEMdroid}, we were unable to obtain the exact value for this metric, but the total average success rate reported in a subsequent work by the same authors~\cite{zhang2024llmbasedabstractionconcretizationgui} indicates a far inferior successful transfer rate of 53\%, which is 44\% lower than \name. 

\subsection{RQ3. \name's Performance and Cost Effectiveness}

With respect to the required time to transfer a test from the source to the target app, \name has significantly better performance than previous methods and transfers each test in 247 seconds on average, which is 290 seconds and 5,120 seconds better than \textsc{TEMdroid} and \textsc{CraftDroid}, respectively. While \textsc{TREADROID} does not report an exact average transfer time, an analysis of the results reported in the paper suggests that it performs faster than \textsc{CraftDroid} but slower than \textsc{TEMdroid}. Therefore, \name outperforms \textsc{TREADROID} on the performance metric as well. 

The main cost of \name is due to the usage of LLMs such as GPT-4o. To calculate the cost for each transfer, we tracked the tokens in each query during the transfer and computed the accumulated cost of all of the queries as the total cost of one transfer. On average, each of the transfers requires 118,600 input tokens and 5,180 output tokens, costing USD \$0.70, which depends on the length of the transferred test. On average, each of the transferred tests has 5.5 steps, and each step costs USD \$0.12.

\subsection{RQ4. Practical Usage of \name on Today’s Popular Apps}

As presented in Section~\ref{Sec_Evaluation}, \name achieved strong results on the \textsc{CraftDroid} dataset, which is the primary benchmark used by existing test transfer techniques. However, since many apps in this dataset are outdated or no longer supported, and given the rapid evolution of mobile app development and workflows, we conducted a follow-up study to evaluate \name on more recent, widely used real-world apps. This study aimed to broaden the evaluation scope and address the dataset's limitations, particularly in categories with fewer functional apps. We selected additional up-to-date apps in the Browser, To Do List, Mail Client and Shopping categories. In each case, we transferred existing \textsc{CraftDroid} tests to the new apps and measured precision, recall, success rate, and reduction, as shown in Table~\ref{table_usefulness}. \name demonstrated strong performance on this new set, achieving an average success rate of 98\%, an average reduction score of 97\%, and an average transfer time of 238 seconds. These findings validate \name's effectiveness in adapting to modern app workflows and highlight the value of transferring tests from older apps to reduce manual testing effort in current app development.
\renewcommand{\arraystretch}{1.1}
\begin{table}[htb]
\centering
\caption{Analysis of precision, recall, success, and reduction metrics achieved by \name, on new apps.}
\resizebox{\textwidth}{!}{
\begin{tabular}{cccccccccc}
\hline
\multirow{2}{*}{\textbf{App}} & \multirow{2}{*}{\textbf{Download}} & \multirow{2}{*}{\textbf{Test}} & \multicolumn{2}{c}{\textbf{Precision}} & \multicolumn{2}{c}{\textbf{Recall}} & \multirow{2}{*}{\textbf{Success}} & \multirow{2}{*}{\textbf{Reduction}} \\ \cline{4-7}
& & & \textbf{GUI Event} & \textbf{Oracle Event} & \textbf{GUI Event} & \textbf{Oracle Event} & & \\  \hline

\multirow{2}{*}{\small{Mozilla Firefox~\cite{firefox}}} & \multirow{2}{*}{100M+} & \small{c1-t1} & 100 & 100 & 100 & 100 & 100 & 100 \\ \hhline{|~|~|-|-|-|-|-|-|-|}
                       &  & \small{c1-t2} & 100 & 100 & 100 & 100 & 100 & 100 \\ \hline

\multirow{2}{*}{\small{ToDo~\cite{splendo}}} & \multirow{2}{*}{10M+} & \small{c2-t1} & 100 & 100 & 100 & 100 & 100 & 100 \\ \hhline{|~|~|-|-|-|-|-|-|-|}
                       &  & \small{c2-t2} & 91.66 & 88.88 & 91.66 & 88.88 & 100 & 90.63 \\ \hline

\multirow{2}{*}{\small{DODuae~\cite{dod}}} & \multirow{2}{*}{1M+} & \small{c3-t1} & 100 & 30 & 100 & 37.5 & 100 & 50 \\ \hhline{|~|~|-|-|-|-|-|-|-|}
                       &  & \small{c3-t2} & 100 & 66.67 & 100 & 66.67 & 100 & 77.78  \\ \hline

\multirow{2}{*}{\small{Zalando~\cite{zalando}}} & \multirow{2}{*}{50M+} & \small{c3-t1} & 94.73 & 40 & 100 & 57.14 & 50 & 62.99\\ \hhline{|~|~|-|-|-|-|-|-|-|}
                       &  & \small{c3-t2} & 81.81 & 66.67 & 100 & 66.67 & 100 & 64.58 \\ \hline

\multirow{2}{*}{\small{Email~\cite{easilydo}}} & \multirow{2}{*}{10M+} & \small{c4-t1} & 100 & 100 & 100 & 100 & 100 & 100\\ \hhline{|~|~|-|-|-|-|-|-|-|}
                       &  & \small{c4-t2} & 100 & 77.78 & 100 & 77.78  & 100 & 84.26 \\ \hline

\end{tabular}
}

\label{table_usefulness}
\end{table}
\vspace{3mm}

\vspace{-20px}
\subsection{RQ5. Evaluating \name on Data Unseen by LLMs}

Since \name leverages existing off-the-shelf LLMs like GPT-4o to transfer tests, it is important to evaluate its effectiveness on apps and test cases that the LLM has not previously encountered. This helps ensure that the promising results discussed earlier are not simply due to the LLM’s prior familiarity with the subjects, and it also validates \name’s applicability to future, unseen apps. To do this, we took two key steps: selecting apps that were released after GPT-4o’s knowledge cutoff date of October 2023~\cite{GPT-4o}, and manually writing new test cases for these apps rather than relying on publicly available ones. We introduced five new app categories and, within each, selected three recently released apps. For each category, we manually designed a representative usage scenario and created corresponding tests, resulting in 30 total test transfers across these new apps.

Table~\ref{table_unseen} outlines the app categories, selected apps, and usage scenarios. Table~\ref{table_RQ5} presents the precision, recall, success rate, reduction, and transfer time metrics for these unseen test transfers. On average, \name achieved 97\% and 89\% precision and 97\% and 92\% recall for GUI and oracle events, respectively. Additionally, it reached a 93\% average successful transfer rate and 88\% reduction score, consistent with the findings in earlier research questions. These outcomes confirm that \name performs reliably on data that was not part of the LLM’s training set, supporting the robustness of the approach.

\renewcommand{\arraystretch}{1.1}
\begin{table}[h]
\centering
\caption{Evaluation subjects used to verify \name's applicability on data previously unseen to GPT-4o.}
\resizebox{\textwidth}{!}{
\begin{tabular}{cccccccccc}
\hline
\multirow{2}{*}
{\textbf{App Category}} & 
\multirow{2}{*}
{\textbf{Subject Apps}} & 
{\textbf{Release}} & 
\multirow{2}{*}
{\textbf{Functionality}} & 
{\textbf{\# Total}} &
{\textbf{\# Oracle}}\\
& &{\textbf{Date}} & &{\textbf{Events}} & {\textbf{Events}}\\ 
\hline

\multirow{3}{*}{\small{c6-EMI Calculator}} & a1-EMI Calculator \& Financial~\cite{emi1} & 9 Aug 2024 & \multirow{3}{*}{\small{t1-Add 3 entries and calculate the EMI}} & \multirow{3}{*}{\small{5.4}} & \multirow{3}{*}{\small{1}}\\ 
\hhline{|~|-|-|~|~|~|}
                       & a2-Cash Loan EMI Calculator~\cite{emi2} & 20 Jan 2025 \\ 
\hhline{|~|-|-|~|~|~|}
                       & a3-EMI Calculator : Loan Planner~\cite{emi3} & 1 Feb 2025 \\ 
\hline

\multirow{3}{*}{\small{c7-AI Chatbots}} & a1-Google Gemini~\cite{chatbo1} & 4 Jun 2024
 & {\small{t1-Start a conversation asking a Yes/No question }} & \multirow{3}{*}{\small{3}} & \multirow{3}{*}{\small{2}}\\ \hhline{|~|-|-|~|~|~|}
                       & a2-Deep Search - AI Chatbot~\cite{chatbo2} & 30 Jan 2025 & \small{and verify the answer}\\
                       \hhline{|~|-|-|~|~|~|}
                       & a3-Chatbot - AI Smart Assistant~\cite{chatbo3} & 16 Feb 2024 \\ 
                       \hline

\multirow{3}{*}{\small{c8-Movies}} & a1-Ava Assistant - Movies \& Shows~\cite{movie1} & 29 Jan 2025 & \multirow{3}{*}{\small{t1-Search for a movie and share the details}} & \multirow{3}{*}{\small{4.5}} & \multirow{3}{*}{\small{1}}\\ \hhline{|~|-|-|~|~|~|}
                       & a2-200TV - Live TV Movies App~\cite{movie2} & 16 Jan 2025 \\
                       \hhline{|~|-|-|~|~|~|}
                       & a3-ClipFix: Movie Shazam~\cite{movie3} & 29 Aug 2024 \\
                       \hline

\multirow{3}{*}{\small{c9-Note}} & a1-Daily Notes - Easy Notebook~\cite{note1} & 1 Feb 2025 & \multirow{3}{*}{\small{t1-Add a note and search for it by its title}} & \multirow{3}{*}{\small{6.5}} & \multirow{3}{*}{\small{2}} \\ \hhline{|~|-|-|~|~|~|}
                       & a2-Notes - QuickNotes~\cite{note2} & 14 Jan 2024 \\ 
                       \hhline{|~|-|-|~|~|~|}
                       &  a3-Personal notes and tasks~\cite{note3} & 5 Jan 2025 \\
                       \hline

\multirow{3}{*}{\small{c10-Messenger}} & a1-Messages for SMS - DUAL SIM~\cite{sms1} & 26 Dec 2024 & {\small{t1-Search for a phone number and send a message}} & \multirow{3}{*}{\small{5}} & \multirow{3}{*}{\small{2}}\\ \hhline{|~|-|-|~|~|~|}
                       &  a2-Messages: Text SMS~\cite{sms2} & 29 Jan 2025 & \small{ to the found recipient}\\ 
                       \hhline{|~|-|-|~|~|~|}
                       &  a3-Color SMS: Message \& Messenger~\cite{sms3} & 26 Oct 2023 \\
                       \hline

\end{tabular}
}

\label{table_unseen}
\end{table}
\vspace{3mm}

\renewcommand{\arraystretch}{1.1}
\begin{table}[htb]
\centering
\caption{Analysis of precision, recall, success, reduction, and transfer time metrics achieved by \name, on unseen apps.}
\resizebox{\textwidth}{!}{
\begin{tabular}{cccccccc}
\hline
\textbf{Category Test} & \multicolumn{2}{c}{\textbf{Precision}} & \multicolumn{2}{c}{\textbf{Recall}} & \textbf{Successful} & \textbf{Reduction} & \textbf{Transfer Time} \\ \cline{2-5}
& \textbf{GUI Event} & \textbf{Oracle Event} & \textbf{GUI Event} & \textbf{Oracle Event} & \textbf{Transfer Rate} & & \\ \hline

\small{c6-t1} & 100 & 100 & 100 & 100 & 100 & 100 & 294.5 \\ \hline
\small{c7-t1} & 100 & 66.7 & 100 & 72.7 & 100 & 60 & 265.44 \\ \hline
\small{c8-t1} & 100 & 30 & 100 & 100 & 100 & 100 & 273.69 \\ \hline
\small{c9-t1} & 91.7 & 91.7 & 91.7 & 91.7 & 66.7 & 83.3 & 333.77 \\ \hline
\small{c10-t1} & 100 & 100 & 100 & 100 & 100 & 100 & 424.63 \\ \hline
\textbf{Average} & \textbf{97.8} & \textbf{89.6} & \textbf{97.8} & \textbf{92.8} & \textbf{93.3} & \textbf{88.6} & \textbf{318.41} \\ \hline

\end{tabular}
}

\label{table_RQ5}
\end{table}
\vspace{3mm}

\section{Discussion}
\label{Sec_Discussion}

Our evaluation demonstrates that \name achieves high accuracy in transferring tests across Android apps, outperforming existing solutions. However, we observed certain failure cases that highlight limitations of the current approach. These include incorrect action selection by the LLM, incomplete flow transfer where critical steps are omitted, and the insufficient oracles generation, particularly for transitions. While these issues do not always prevent test execution, they can compromise the correctness or completeness of the transferred tests. Readers can find specific examples of these cases on the project repository~\cite{repo}.

Additionally, \name shares some common limitations with prior test transfer techniques. Its performance is influenced by the accessibility and quality of UI metadata in the target app, particularly when meaningful attributes like \texttt{resource-id} or \texttt{content-desc} are missing. The tool also struggles with transient UI elements such as toast messages, which may disappear before they can be processed. Furthermore, certain events, like user registration, can be irreversible, making it difficult to recover from partial failures. These challenges suggest directions for future enhancements, such as incorporating visual analysis~\cite{chen2020unblind} or improving app state management.

\section{Threats to Validity}
\label{Sec_Threat}

An important threat to the validity of our work stems from our reliance on off-the-shelf LLMs such as GPT-4o. These models produce responses that are not fully deterministic. Consequently, our approach may also produce varying results for the same test, impacting the findings' reproducibility. We tried to address this threat by querying the LLM multiple times, as discussed in Section~\ref{sec:test-migration}. 

Another threat to the validity of our evaluation comes from our inability to run the previous test transfer tools. This is due to reasons such as the unavailability of parts of the source code~\cite{temdroid} and technical problems that are mainly due to changes and the lack of maintenance of existing dependencies~\cite{lin2019craftdroid,liu2024enhancing}. As detailed in Section~\ref{Sec_Evaluation}, to navigate this issue and offer a comparative analysis, we utilized the publicly available \textsc{CraftDroid} dataset~\cite{lin2019craftdroid} previously used by existing tools. A related threat arises from a small number of apps from the \textsc{CraftDroid} dataset~\cite{lin2019craftdroid} that are deprecated. To mitigate this threat and ensure a fair comparison, for \textsc{CraftDroid}~\cite{lin2019craftdroid} and \textsc{TREADROID}~\cite{liu2024enhancing}, which provided detailed evaluation results for individual transfers, we restricted our comparison to the subset of apps from the dataset that are still functional. As discussed in Section~\ref{Sec_Evaluation}, for \textsc{TEMdroid}~\cite{temdroid}, we could not obtain detailed experimental results even after contacting the authors. Due to the lack of available data, with respect to \textsc{TEMdroid}, we compared the average numbers for each category, which can be a threat to validity.

Our tool leverages Appium and the Appium UiAutomator2 Driver~\cite{uiautomator} to interact with UI elements. This driver follows the WebDriver standard~\cite{webdriver}, enabling a wide range of interactions with the app UI. However, since our tool builds on earlier approaches, we adopt a similarly limited yet well-curated set of interactions. While we support all actions covered by previous approaches—enhancing selector accuracy in the process—we do not provide full support for every possible action.
\section{Related Works}
\label{Sec_Related}
\looseness=-1

The most relevant group of works targets transferring tests from one Android app to another, similar to \name. Behrang et al.~\cite{behrang2018test, behrang2019atm} and Lin et al.~\cite{lin2019craftdroid} proposed \textsc{ATM} and \textsc{CraftDroid}, which rely on app analysis and NLP techniques to transfer tests across different Android apps within the same domain.  \textsc{AppFlow}~\cite{hu2018appflow} is a machine learning-based approach that utilizes screen and widget classification to generate UI tests for an app using a library of existing tests. Liu et al.~\cite{liu2022test, liu2024enhancing} proposed adaptive semantic matching strategies for test transfer. Yu et al.~\cite{temdroid} have recently proposed \textsc{TEMdroid},
a semantic matching-based approach for test transfer that leverages dynamic analysis and Siamese networks. Zhao et al.~\cite{Zhao_2020} proposed \textsc{FrUITeR}, a framework for automatically evaluating the previous test transfer approaches. Mariani et al.~\cite{mariani2021evolutionary} presented a study on techniques for semantic matching of GUI events used by existing test reuse approaches. Khalili et al. proposed \textsc{SemFinder}~\cite{khalili2024semantic}, an
approach that assesses different configurations for mapping UI events across apps based on
their textual information but does not focus on the test transfer problem as a whole. Mishra
et al.~\cite{mishra2023image} extended \textsc{SemFinder} by incorporating visual information into the event mapping technique. 

Most of the techniques mentioned above that directly target test transfer rely on similarity-based matching between the events of source and target apps. These approaches can be highly effective in cases where the source and target apps have similar workflows for the functionality under test. However,
as research work showed~\cite{Zhao_2020}, in practice, due to the inherent differences between apps even
within the same domain, relying solely on event-by-event similarity-based matching may not
result in useful transferred tests. In contrast with these techniques, \name addresses the test transfer holistically and does not depend on event-by-event similarity-based matching.

Another group of recent works targets a slightly different problem, focusing on many-to-one UI test transfers. \textsc{MigratePro}~\cite{zhang2024synthesis} is a technique that aims to improve UI test transfer by generating a new test from multiple tests that have already been migrated to the target app from various source apps. \textsc{MigratePro} is not a transfer technique itself, and it improves tests transferred by an existing transfer technique. Future research can explore using tests transferred by \name as input for \textsc{MigratePro} to assess potential improvements. Another recent work, \textsc{MACdroid}~\cite{zhang2024llmbasedabstractionconcretizationgui}, uses LLMs to create tests for a new target app using an abstract test logic created from multiple source tests targeting the same functionality on different apps. Note that compared to one-to-one transfer, many-to-one transfer is a less challenging problem due to the availability of more data, such as multiple different flows in various apps that can be more similar to the intended flow in the target app. However, in practice, there often are not multiple compatible tests for the same functionality available that can be used for transfer, which limits the practicality of these techniques. Furthermore, unlike \name, which employs multimodal LLMs and utilizes both visual and textual information for UI understanding tasks, \textsc{MACdroid} relies solely on textual data, which can lead to certain limitations, as discussed in Section~\ref{Sec_Approach}. A more detailed comparison of the two approaches is not possible due to the unavailability of the implementation and artifacts relevant to this technique at the time of publication.

There exists another group of relevant research work that targets transferring UI tests across different platforms. \textsc{TestMig}~\cite{ISSTA19TestMig} and \textsc{MAPIT}~\cite{mapit} have targeted test transfer across Android and iOS apps. Ji et al.~\cite{ji2023vision} conducted a comprehensive study on
vision-based widget mapping for cross-platform GUI test migration. In the context of web apps, Rau et al.~\cite{rau2018transferring} proposed an approach for efficiently generating UI tests by learning from the existing tests of other apps. Mariani et al.~\cite{mariani2018augusto} proposed an approach that automatically exploits the common functionalities of Java apps to generate UI tests. \textsc{TransDroid} \cite{lin2022gui} has transferred tests from a web app to its Android version by making use of a navigation graph and the textual data of the events and widgets involved in them. \textsc{MUT}~\cite{gao2024mut} is a technique for transferring GUI tests of one web app to another using NLP methods. 

Another group of related research focuses on bug reproduction in Android apps~\cite{wang2024feedback, zhao2019recdroid, zhao2019automatically}. Note that although these works and test migration efforts both aim to execute a sequence of events in an Android app, the problems differ in two important aspects: (1) One of the biggest challenges of test migration comes from the differences in the flow of executed steps for a scenario between the source and target app. However, this issue does not exist in bug reproduction since the bug report belongs to the same app. (2) Test transfer also involves the challenge of accurately transferring and creating oracle events, which is a complexity that is not a part of the bug reproduction task.

Finally, several recent publications~\cite{kang2023evaluating, kang2023large, feng2024prompting, huang2024crashtranslator, liu2024make, liu2024vision, liu2023fill, yu2023llm, yoon2023autonomous, wen2023droidbot, ju2024study} have explored the application of LLMs to advance mobile testing. However, none of these techniques address the automated transfer of existing usage-based tests across apps with similar functionality. 
\section{Conclusion and Future Work}\label{Sec_Conclusion}
\looseness=-1
This paper has presented \name, a technique that relies on multimodal LLMs for transferring usage-based UI tests across Android apps. In our extensive evaluation covering five app categories, \name was able to successfully transfer 97\% of tests and reduce more than 90\% of the total manual work required for writing UI tests.

Potential future areas of work can target expanding our technique, particularly for development across various platforms such as Web and iOS, which promises to yield significant time savings in end-to-end test development and maintenance. Another possible research direction is the study of how integrating the approaches~\cite{chen2020unblind} that enhance app accessibility can improve the applicability of techniques such as \name on non-accessible apps.
\section{Data Availability}
\label{Sec_Data_Availability}

\name's implementation and all of our research artifacts are available publicly~\cite{repo}. 
\section{Acknowledgment}
\label{Sec_Acknowledgment}

This work has been supported, in part, by award numbers 2106871, 2106306, and 2211790 from the U.S. National Science Foundation.

\vspace{10pt}

\bibliographystyle{ACM-Reference-Format}
\bibliography{ts}


\begin{thebibliography}{104}

%%% the Web address.

\ifx \showCODEN    \undefined \def \showCODEN     #1{\unskip}     \fi
\ifx \showDOI      \undefined \def \showDOI       #1{#1}\fi
\ifx \showISBNx    \undefined \def \showISBNx     #1{\unskip}     \fi
\ifx \showISBNxiii \undefined \def \showISBNxiii  #1{\unskip}     \fi
\ifx \showISSN     \undefined \def \showISSN      #1{\unskip}     \fi
\ifx \showLCCN     \undefined \def \showLCCN      #1{\unskip}     \fi
\ifx \shownote     \undefined \def \shownote      #1{#1}          \fi
\ifx \showarticletitle \undefined \def \showarticletitle #1{#1}   \fi
\ifx \showURL      \undefined \def \showURL       {\relax}        \fi
% invisible to TeX
\providecommand\bibfield[2]{#2}
\providecommand\bibinfo[2]{#2}
\providecommand\natexlab[1]{#1}
\providecommand\showeprint[2][]{arXiv:#2}

\bibitem[dod(2024)]%
        {dod}
 \bibinfo{year}{2024}\natexlab{}.
\newblock \bibinfo{title}{DODuae - Women's Online Store}.
\newblock \bibinfo{howpublished}{\url{https://tinyurl.com/mu5zkenz}}.
\newblock


\bibitem[eas(2024)]%
        {easilydo}
 \bibinfo{year}{2024}\natexlab{}.
\newblock \bibinfo{title}{Email - Fast \& Secure Mail}.
\newblock \bibinfo{howpublished}{\url{https://tinyurl.com/53ensprk}}.
\newblock


\bibitem[fir(2024)]%
        {firefox}
 \bibinfo{year}{2024}\natexlab{}.
\newblock \bibinfo{title}{Firefox Fast \& Private Browser}.
\newblock \bibinfo{howpublished}{\url{https://tinyurl.com/yc5t5tkh}}.
\newblock


\bibitem[spl(2024)]%
        {splendo}
 \bibinfo{year}{2024}\natexlab{}.
\newblock \bibinfo{title}{To Do List}.
\newblock \bibinfo{howpublished}{\url{https://tinyurl.com/c8chz4fb}}.
\newblock


\bibitem[zal(2024)]%
        {zalando}
 \bibinfo{year}{2024}\natexlab{}.
\newblock \bibinfo{title}{Zalando – Online Fashion Store}.
\newblock \bibinfo{howpublished}{\url{https://tinyurl.com/mpee366u}}.
\newblock


\bibitem[mov(2025a)]%
        {movie2}
 \bibinfo{year}{2025}\natexlab{a}.
\newblock \bibinfo{title}{200TV - Live TV Movies App}.
\newblock \bibinfo{howpublished}{\url{https://tinyurl.com/3xf6wb8v}}.
\newblock


\bibitem[mov(2025b)]%
        {movie1}
 \bibinfo{year}{2025}\natexlab{b}.
\newblock \bibinfo{title}{Ava Assistant - Movies \& Shows}.
\newblock \bibinfo{howpublished}{\url{https://tinyurl.com/4e2pzxry}}.
\newblock


\bibitem[emi(2025a)]%
        {emi2}
 \bibinfo{year}{2025}\natexlab{a}.
\newblock \bibinfo{title}{Cash Loan EMI Calcualtor}.
\newblock \bibinfo{howpublished}{\url{https://tinyurl.com/2tb59nr7}}.
\newblock


\bibitem[cha(2025a)]%
        {chatbo3}
 \bibinfo{year}{2025}\natexlab{a}.
\newblock \bibinfo{title}{Chatbot - AI Smart Assistant}.
\newblock \bibinfo{howpublished}{\url{https://tinyurl.com/vk4t739r}}.
\newblock


\bibitem[mov(2025c)]%
        {movie3}
 \bibinfo{year}{2025}\natexlab{c}.
\newblock \bibinfo{title}{ClipFix: Movie Shazam}.
\newblock \bibinfo{howpublished}{\url{https://tinyurl.com/avjp9bzv}}.
\newblock


\bibitem[sms(2025a)]%
        {sms3}
 \bibinfo{year}{2025}\natexlab{a}.
\newblock \bibinfo{title}{Color SMS: Message \& Messenger}.
\newblock \bibinfo{howpublished}{\url{https://tinyurl.com/428xy74b}}.
\newblock


\bibitem[not(2025a)]%
        {note1}
 \bibinfo{year}{2025}\natexlab{a}.
\newblock \bibinfo{title}{Daily Notes - Easy Notebook}.
\newblock \bibinfo{howpublished}{\url{https://tinyurl.com/rrr9j5ee}}.
\newblock


\bibitem[cha(2025b)]%
        {chatbo2}
 \bibinfo{year}{2025}\natexlab{b}.
\newblock \bibinfo{title}{Deep Search - AI Chatbot}.
\newblock \bibinfo{howpublished}{\url{https://tinyurl.com/5c68xzdt}}.
\newblock


\bibitem[emi(2025b)]%
        {emi1}
 \bibinfo{year}{2025}\natexlab{b}.
\newblock \bibinfo{title}{EMI Calculator \& Financial}.
\newblock \bibinfo{howpublished}{\url{https://tinyurl.com/2kbtpd25}}.
\newblock


\bibitem[emi(2025c)]%
        {emi3}
 \bibinfo{year}{2025}\natexlab{c}.
\newblock \bibinfo{title}{EMI Calculator : Loan Planner}.
\newblock \bibinfo{howpublished}{\url{https://tinyurl.com/2vsb3h3y}}.
\newblock


\bibitem[cha(2025c)]%
        {chatbo1}
 \bibinfo{year}{2025}\natexlab{c}.
\newblock \bibinfo{title}{Google Gemini}.
\newblock \bibinfo{howpublished}{\url{https://tinyurl.com/nhe8hpty}}.
\newblock


\bibitem[rep(2025)]%
        {repo}
 \bibinfo{year}{2025}\natexlab{}.
\newblock \bibinfo{title}{LLMigrate open-source repository}.
\newblock \bibinfo{howpublished}{\url{https://github.com/seal-hub/llmigrate}}.
\newblock


\bibitem[sms(2025b)]%
        {sms1}
 \bibinfo{year}{2025}\natexlab{b}.
\newblock \bibinfo{title}{Messages for SMS - DUAL SIM}.
\newblock \bibinfo{howpublished}{\url{https://tinyurl.com/mrd2rdus}}.
\newblock


\bibitem[sms(2025c)]%
        {sms2}
 \bibinfo{year}{2025}\natexlab{c}.
\newblock \bibinfo{title}{Messages: Text SMS}.
\newblock \bibinfo{howpublished}{\url{https://tinyurl.com/bdfhk5xk}}.
\newblock


\bibitem[not(2025b)]%
        {note2}
 \bibinfo{year}{2025}\natexlab{b}.
\newblock \bibinfo{title}{Notes - QuickNotes}.
\newblock \bibinfo{howpublished}{\url{https://tinyurl.com/mue2jyau}}.
\newblock


\bibitem[not(2025c)]%
        {note3}
 \bibinfo{year}{2025}\natexlab{c}.
\newblock \bibinfo{title}{Personal notes and tasks}.
\newblock \bibinfo{howpublished}{\url{https://tinyurl.com/393fn3ww}}.
\newblock


\bibitem[Ahmed and Devanbu(2022)]%
        {ahmed2022few}
\bibfield{author}{\bibinfo{person}{Toufique Ahmed} {and} \bibinfo{person}{Premkumar Devanbu}.} \bibinfo{year}{2022}\natexlab{}.
\newblock \showarticletitle{Few-shot training LLMs for project-specific code-summarization}. In \bibinfo{booktitle}{\emph{Proceedings of the 37th IEEE/ACM International Conference on Automated Software Engineering}}. \bibinfo{pages}{1--5}.
\newblock


\bibitem[Amalfitano et~al\mbox{.}(2012)]%
        {amalfitano2012using}
\bibfield{author}{\bibinfo{person}{Domenico Amalfitano}, \bibinfo{person}{Anna~Rita Fasolino}, \bibinfo{person}{Porfirio Tramontana}, \bibinfo{person}{Salvatore De~Carmine}, {and} \bibinfo{person}{Atif~M Memon}.} \bibinfo{year}{2012}\natexlab{}.
\newblock \showarticletitle{Using GUI ripping for automated testing of Android applications}. In \bibinfo{booktitle}{\emph{2012 Proceedings of the 27th IEEE/ACM International Conference on Automated Software Engineering}}. IEEE, \bibinfo{pages}{258--261}.
\newblock


\bibitem[Amalfitano et~al\mbox{.}(2015)]%
        {MobiGuitar}
\bibfield{author}{\bibinfo{person}{D. Amalfitano}, \bibinfo{person}{A.~R. Fasolino}, \bibinfo{person}{P. Tramontana}, \bibinfo{person}{B.~D. Ta}, {and} \bibinfo{person}{A.~M. Memon}.} \bibinfo{year}{2015}\natexlab{}.
\newblock \showarticletitle{MobiGUITAR: Automated Model-Based Testing of Mobile Apps}.
\newblock \bibinfo{journal}{\emph{IEEE Software}} \bibinfo{volume}{32}, \bibinfo{number}{5} (\bibinfo{date}{Sept} \bibinfo{year}{2015}), \bibinfo{pages}{53--59}.
\newblock
\showISSN{0740-7459}
\urldef\tempurl%
\url{https://doi.org/10.1109/MS.2014.55}
\showDOI{\tempurl}


\bibitem[Anand et~al\mbox{.}(2012)]%
        {FSE12Concolic}
\bibfield{author}{\bibinfo{person}{Saswat Anand}, \bibinfo{person}{Mayur Naik}, \bibinfo{person}{Mary~Jean Harrold}, {and} \bibinfo{person}{Hongseok Yang}.} \bibinfo{year}{2012}\natexlab{}.
\newblock \showarticletitle{Automated Concolic Testing of Smartphone Apps}. In \bibinfo{booktitle}{\emph{Proceedings of the ACM SIGSOFT 20th International Symposium on the Foundations of Software Engineering}} (Cary, North Carolina) \emph{(\bibinfo{series}{FSE '12})}. \bibinfo{publisher}{ACM}, \bibinfo{address}{New York, NY, USA}, Article \bibinfo{articleno}{59}, \bibinfo{numpages}{11}~pages.
\newblock
\showISBNx{978-1-4503-1614-9}
\urldef\tempurl%
\url{https://doi.org/10.1145/2393596.2393666}
\showDOI{\tempurl}


\bibitem[Behrang and Orso(2018)]%
        {behrang2018test}
\bibfield{author}{\bibinfo{person}{Farnaz Behrang} {and} \bibinfo{person}{Alessandro Orso}.} \bibinfo{year}{2018}\natexlab{}.
\newblock \showarticletitle{Test migration for efficient large-scale assessment of mobile app coding assignments}. In \bibinfo{booktitle}{\emph{Proceedings of the 27th ACM SIGSOFT International Symposium on Software Testing and Analysis}}.
\newblock


\bibitem[Behrang and Orso(2019)]%
        {behrang2019atm}
\bibfield{author}{\bibinfo{person}{Farnaz Behrang} {and} \bibinfo{person}{Alessandro Orso}.} \bibinfo{year}{2019}\natexlab{}.
\newblock \showarticletitle{Test Migration Between Mobile Apps with Similar Functionality}. In \bibinfo{booktitle}{\emph{34th International Conference on Automated Software Engineering (ASE 2019)}}.
\newblock


\bibitem[Behrang and Orso(pear)]%
        {ASE19Orso}
\bibfield{author}{\bibinfo{person}{Farnaz Behrang} {and} \bibinfo{person}{Alessandro Orso}.} \bibinfo{year}{2019. To appear.}\natexlab{}.
\newblock \showarticletitle{Test Migration Between Mobile Apps with Similar Functionality}. In \bibinfo{booktitle}{\emph{Proceedings of the The 34th IEEE/ACM International Conference on Automated Software Engineering}} (San Diego, USA) \emph{(\bibinfo{series}{ASE '19})}.
\newblock


\bibitem[Bouzenia et~al\mbox{.}(2024)]%
        {bouzenia2024repairagent}
\bibfield{author}{\bibinfo{person}{Islem Bouzenia}, \bibinfo{person}{Premkumar Devanbu}, {and} \bibinfo{person}{Michael Pradel}.} \bibinfo{year}{2024}\natexlab{}.
\newblock \showarticletitle{Repairagent: An autonomous, llm-based agent for program repair}.
\newblock \bibinfo{journal}{\emph{arXiv preprint arXiv:2403.17134}} (\bibinfo{year}{2024}).
\newblock


\bibitem[Chen et~al\mbox{.}(2020)]%
        {chen2020unblind}
\bibfield{author}{\bibinfo{person}{Jieshan Chen}, \bibinfo{person}{Chunyang Chen}, \bibinfo{person}{Zhenchang Xing}, \bibinfo{person}{Xiwei Xu}, \bibinfo{person}{Liming Zhu}, \bibinfo{person}{Guoqiang Li}, {and} \bibinfo{person}{Jinshui Wang}.} \bibinfo{year}{2020}\natexlab{}.
\newblock \showarticletitle{Unblind your apps: Predicting natural-language labels for mobile gui components by deep learning}. In \bibinfo{booktitle}{\emph{Proceedings of the ACM/IEEE 42nd international conference on software engineering}}. \bibinfo{pages}{322--334}.
\newblock


\bibitem[Choi et~al\mbox{.}(2013)]%
        {OOPLSA13MinimalRestart}
\bibfield{author}{\bibinfo{person}{Wontae Choi}, \bibinfo{person}{George Necula}, {and} \bibinfo{person}{Koushik Sen}.} \bibinfo{year}{2013}\natexlab{}.
\newblock \showarticletitle{Guided GUI Testing of Android Apps with Minimal Restart and Approximate Learning}. In \bibinfo{booktitle}{\emph{Proceedings of the 2013 ACM SIGPLAN International Conference on Object Oriented Programming Systems Languages \& Applications}} (Indianapolis, Indiana, USA) \emph{(\bibinfo{series}{OOPSLA '13})}. \bibinfo{publisher}{ACM}, \bibinfo{address}{New York, NY, USA}, \bibinfo{pages}{623--640}.
\newblock
\showISBNx{978-1-4503-2374-1}
\urldef\tempurl%
\url{https://doi.org/10.1145/2509136.2509552}
\showDOI{\tempurl}


\bibitem[Consortium(2025)]%
        {webdriver}
\bibfield{author}{\bibinfo{person}{World Wide~Web Consortium}.} \bibinfo{year}{2025}\natexlab{}.
\newblock \bibinfo{howpublished}{\url{https://www.w3.org/TR/webdriver/}}.
\newblock


\bibitem[Contributors({[n.\,d.]})]%
        {appium}
\bibfield{author}{\bibinfo{person}{Appium Contributors}.} \bibinfo{year}{[n.\,d.]}\natexlab{}.
\newblock \bibinfo{title}{Appium}.
\newblock \bibinfo{howpublished}{\url{https://github.com/appium/appium}}.
\newblock


\bibitem[Deng et~al\mbox{.}(2024)]%
        {deng2024mind2web}
\bibfield{author}{\bibinfo{person}{Xiang Deng} {et~al\mbox{.}}} \bibinfo{year}{2024}\natexlab{}.
\newblock \showarticletitle{Mind2web: Towards a generalist agent for the web}.
\newblock \bibinfo{journal}{\emph{Advances in Neural Information Processing Systems}}  \bibinfo{volume}{36} (\bibinfo{year}{2024}).
\newblock


\bibitem[Dong et~al\mbox{.}(2020)]%
        {dong2020time}
\bibfield{author}{\bibinfo{person}{Zhen Dong}, \bibinfo{person}{Marcel B{\"o}hme}, \bibinfo{person}{Lucia Cojocaru}, {and} \bibinfo{person}{Abhik Roychoudhury}.} \bibinfo{year}{2020}\natexlab{}.
\newblock \showarticletitle{Time-travel testing of Android apps}. In \bibinfo{booktitle}{\emph{2020 IEEE/ACM 42nd International Conference on Software Engineering (ICSE)}}. IEEE, \bibinfo{pages}{481--492}.
\newblock


\bibitem[Ermuth and Pradel(2016)]%
        {ISSTA16MonkeySee}
\bibfield{author}{\bibinfo{person}{Markus Ermuth} {and} \bibinfo{person}{Michael Pradel}.} \bibinfo{year}{2016}\natexlab{}.
\newblock \showarticletitle{Monkey See, Monkey Do: Effective Generation of GUI Tests with Inferred Macro Events}. In \bibinfo{booktitle}{\emph{Proceedings of the 25th International Symposium on Software Testing and Analysis}} (Saarbr\&\#252;cken, Germany) \emph{(\bibinfo{series}{ISSTA 2016})}. \bibinfo{publisher}{ACM}, \bibinfo{address}{New York, NY, USA}, \bibinfo{pages}{82--93}.
\newblock
\showISBNx{978-1-4503-4390-9}
\urldef\tempurl%
\url{https://doi.org/10.1145/2931037.2931053}
\showDOI{\tempurl}


\bibitem[Feng and Chen(2024)]%
        {feng2024prompting}
\bibfield{author}{\bibinfo{person}{Sidong Feng} {and} \bibinfo{person}{Chunyang Chen}.} \bibinfo{year}{2024}\natexlab{}.
\newblock \showarticletitle{Prompting is all you need: Automated android bug replay with large language models}. In \bibinfo{booktitle}{\emph{Proceedings of the 46th IEEE/ACM International Conference on Software Engineering}}. \bibinfo{pages}{1--13}.
\newblock


\bibitem[Gao et~al\mbox{.}(2024)]%
        {gao2024mut}
\bibfield{author}{\bibinfo{person}{Yi Gao}, \bibinfo{person}{Xing Hu}, \bibinfo{person}{Tongtong Xu}, \bibinfo{person}{Xin Xia}, \bibinfo{person}{David Lo}, {and} \bibinfo{person}{Xiaohu Yang}.} \bibinfo{year}{2024}\natexlab{}.
\newblock \showarticletitle{MUT: Human-in-the-Loop Unit Test Migration}. In \bibinfo{booktitle}{\emph{Proceedings of the IEEE/ACM 46th International Conference on Software Engineering}}. \bibinfo{pages}{1--12}.
\newblock


\bibitem[Gu et~al\mbox{.}(2019)]%
        {gu2019practicalAPE}
\bibfield{author}{\bibinfo{person}{Tianxiao Gu} {et~al\mbox{.}}} \bibinfo{year}{2019}\natexlab{}.
\newblock \showarticletitle{Practical GUI testing of Android applications via model abstraction and refinement}. In \bibinfo{booktitle}{\emph{2019 IEEE/ACM 41st International Conference on Software Engineering (ICSE)}}. IEEE, \bibinfo{pages}{269--280}.
\newblock


\bibitem[Gur et~al\mbox{.}(2022)]%
        {gur2022understanding}
\bibfield{author}{\bibinfo{person}{Izzeddin Gur} {et~al\mbox{.}}} \bibinfo{year}{2022}\natexlab{}.
\newblock \showarticletitle{Understanding html with large language models}.
\newblock \bibinfo{journal}{\emph{arXiv preprint arXiv:2210.03945}} (\bibinfo{year}{2022}).
\newblock


\bibitem[Gur et~al\mbox{.}(2023)]%
        {gur2023real}
\bibfield{author}{\bibinfo{person}{Izzeddin Gur} {et~al\mbox{.}}} \bibinfo{year}{2023}\natexlab{}.
\newblock \showarticletitle{A real-world webagent with planning, long context understanding, and program synthesis}.
\newblock \bibinfo{journal}{\emph{arXiv preprint arXiv:2307.12856}} (\bibinfo{year}{2023}).
\newblock


\bibitem[Haas et~al\mbox{.}(2021)]%
        {haas2021manual}
\bibfield{author}{\bibinfo{person}{Roman Haas} {et~al\mbox{.}}} \bibinfo{year}{2021}\natexlab{}.
\newblock \showarticletitle{How can manual testing processes be optimized? developer survey, optimization guidelines, and case studies}. In \bibinfo{booktitle}{\emph{Proceedings of the 29th ACM Joint Meeting on European Software Engineering Conference and Symposium on the Foundations of Software Engineering}}. \bibinfo{pages}{1281--1291}.
\newblock


\bibitem[Hao et~al\mbox{.}(2014)]%
        {PUMA}
\bibfield{author}{\bibinfo{person}{Shuai Hao}, \bibinfo{person}{Bin Liu}, \bibinfo{person}{Suman Nath}, \bibinfo{person}{William~G.J. Halfond}, {and} \bibinfo{person}{Ramesh Govindan}.} \bibinfo{year}{2014}\natexlab{}.
\newblock \showarticletitle{PUMA: Programmable UI-automation for Large-scale Dynamic Analysis of Mobile Apps}. In \bibinfo{booktitle}{\emph{Proceedings of the 12th Annual International Conference on Mobile Systems, Applications, and Services}} (Bretton Woods, New Hampshire, USA) \emph{(\bibinfo{series}{MobiSys '14})}. \bibinfo{publisher}{ACM}, \bibinfo{address}{New York, NY, USA}, \bibinfo{pages}{204--217}.
\newblock
\showISBNx{978-1-4503-2793-0}
\urldef\tempurl%
\url{https://doi.org/10.1145/2594368.2594390}
\showDOI{\tempurl}


\bibitem[Hu et~al\mbox{.}(2018)]%
        {hu2018appflow}
\bibfield{author}{\bibinfo{person}{Gang Hu}, \bibinfo{person}{Linjie Zhu}, {and} \bibinfo{person}{Junfeng Yang}.} \bibinfo{year}{2018}\natexlab{}.
\newblock \showarticletitle{AppFlow: using machine learning to synthesize robust, reusable UI tests}. In \bibinfo{booktitle}{\emph{Proceedings of the 2018 26th ACM Joint Meeting on European Software Engineering Conference and Symposium on the Foundations of Software Engineering}}. ACM, \bibinfo{pages}{269--282}.
\newblock


\bibitem[Huang et~al\mbox{.}(2024)]%
        {huang2024crashtranslator}
\bibfield{author}{\bibinfo{person}{Yuchao Huang} {et~al\mbox{.}}} \bibinfo{year}{2024}\natexlab{}.
\newblock \showarticletitle{Crashtranslator: Automatically reproducing mobile application crashes directly from stack trace}. In \bibinfo{booktitle}{\emph{Proceedings of the 46th IEEE/ACM International Conference on Software Engineering}}. \bibinfo{pages}{1--13}.
\newblock


\bibitem[Jensen et~al\mbox{.}(2013)]%
        {ISSTA13TargetedEvent}
\bibfield{author}{\bibinfo{person}{Casper~S. Jensen}, \bibinfo{person}{Mukul~R. Prasad}, {and} \bibinfo{person}{Anders M{\o}ller}.} \bibinfo{year}{2013}\natexlab{}.
\newblock \showarticletitle{Automated Testing with Targeted Event Sequence Generation}. In \bibinfo{booktitle}{\emph{Proceedings of the 2013 International Symposium on Software Testing and Analysis}} (Lugano, Switzerland) \emph{(\bibinfo{series}{ISSTA 2013})}. \bibinfo{publisher}{ACM}, \bibinfo{address}{New York, NY, USA}, \bibinfo{pages}{67--77}.
\newblock
\showISBNx{978-1-4503-2159-4}
\urldef\tempurl%
\url{https://doi.org/10.1145/2483760.2483777}
\showDOI{\tempurl}


\bibitem[Ji et~al\mbox{.}(2023)]%
        {ji2023vision}
\bibfield{author}{\bibinfo{person}{Ruihua Ji} {et~al\mbox{.}}} \bibinfo{year}{2023}\natexlab{}.
\newblock \showarticletitle{Vision-Based Widget Mapping for Test Migration Across Mobile Platforms: Are We There Yet?}. In \bibinfo{booktitle}{\emph{2023 38th IEEE/ACM International Conference on Automated Software Engineering (ASE)}}. IEEE, \bibinfo{pages}{1416--1428}.
\newblock


\bibitem[Ju et~al\mbox{.}(2024)]%
        {ju2024study}
\bibfield{author}{\bibinfo{person}{Bangyan Ju} {et~al\mbox{.}}} \bibinfo{year}{2024}\natexlab{}.
\newblock \showarticletitle{A Study of Using Multimodal LLMs for Non-Crash Functional Bug Detection in Android Apps}.
\newblock \bibinfo{journal}{\emph{arXiv preprint arXiv:2407.19053}} (\bibinfo{year}{2024}).
\newblock


\bibitem[Kaasila et~al\mbox{.}(2012)]%
        {kaasila2012testdroid}
\bibfield{author}{\bibinfo{person}{Jouko Kaasila}, \bibinfo{person}{Denzil Ferreira}, \bibinfo{person}{Vassilis Kostakos}, {and} \bibinfo{person}{Timo Ojala}.} \bibinfo{year}{2012}\natexlab{}.
\newblock \showarticletitle{Testdroid: automated remote UI testing on Android}. In \bibinfo{booktitle}{\emph{Proceedings of the 11th International Conference on Mobile and Ubiquitous Multimedia}}. \bibinfo{pages}{1--4}.
\newblock


\bibitem[Kang et~al\mbox{.}(2023b)]%
        {kang2023evaluating}
\bibfield{author}{\bibinfo{person}{Sungmin Kang}, \bibinfo{person}{Juyeon Yoon}, \bibinfo{person}{Nargiz Askarbekkyzy}, {and} \bibinfo{person}{Shin Yoo}.} \bibinfo{year}{2023}\natexlab{b}.
\newblock \showarticletitle{Evaluating Diverse Large Language Models for Automatic and General Bug Reproduction}.
\newblock \bibinfo{journal}{\emph{arXiv preprint arXiv:2311.04532}} (\bibinfo{year}{2023}).
\newblock


\bibitem[Kang et~al\mbox{.}(2023a)]%
        {kang2023large}
\bibfield{author}{\bibinfo{person}{Sungmin Kang}, \bibinfo{person}{Juyeon Yoon}, {and} \bibinfo{person}{Shin Yoo}.} \bibinfo{year}{2023}\natexlab{a}.
\newblock \showarticletitle{Large language models are few-shot testers: Exploring llm-based general bug reproduction}. In \bibinfo{booktitle}{\emph{2023 IEEE/ACM 45th International Conference on Software Engineering (ICSE)}}. IEEE, \bibinfo{pages}{2312--2323}.
\newblock


\bibitem[Khalili et~al\mbox{.}(2024)]%
        {khalili2024semantic}
\bibfield{author}{\bibinfo{person}{Farideh Khalili}, \bibinfo{person}{Leonardo Mariani}, \bibinfo{person}{Ali Mohebbi}, \bibinfo{person}{Mauro Pezz{\`e}}, {and} \bibinfo{person}{Valerio Terragni}.} \bibinfo{year}{2024}\natexlab{}.
\newblock \showarticletitle{Semantic matching in GUI test reuse}.
\newblock \bibinfo{journal}{\emph{Empirical Software Engineering}} \bibinfo{volume}{29}, \bibinfo{number}{3} (\bibinfo{year}{2024}), \bibinfo{pages}{1--58}.
\newblock


\bibitem[Kochhar et~al\mbox{.}(2015)]%
        {Kochar2015Study}
\bibfield{author}{\bibinfo{person}{Pavneet~Singh Kochhar} {et~al\mbox{.}}} \bibinfo{year}{2015}\natexlab{}.
\newblock \showarticletitle{Understanding the Test Automation Culture of App Developers}. In \bibinfo{booktitle}{\emph{2015 IEEE 8th International Conference on Software Testing, Verification and Validation (ICST)}}. \bibinfo{pages}{1--10}.
\newblock
\urldef\tempurl%
\url{https://doi.org/10.1109/ICST.2015.7102609}
\showDOI{\tempurl}


\bibitem[Koroglu et~al\mbox{.}(2018)]%
        {ICST18QBE}
\bibfield{author}{\bibinfo{person}{Yavuz Koroglu} {et~al\mbox{.}}} \bibinfo{year}{2018}\natexlab{}.
\newblock \showarticletitle{QBE: QLearning-based exploration of android applications}. In \bibinfo{booktitle}{\emph{Software Testing, Verification and Validation (ICST), 2018 IEEE 11th International Conference on}}. IEEE, \bibinfo{pages}{105--115}.
\newblock


\bibitem[Lab(2024)]%
        {FirebaseRobo}
\bibfield{author}{\bibinfo{person}{Firebase~Test Lab}.} \bibinfo{year}{2024}\natexlab{}.
\newblock \bibinfo{title}{Robo test (Android)}.
\newblock \bibinfo{howpublished}{\url{https://firebase.google.com/docs/test-lab/android/robo-ux-test}}.
\newblock


\bibitem[Levenshtein(1966)]%
        {levenshtein1966binary}
\bibfield{author}{\bibinfo{person}{Vladimir~I Levenshtein}.} \bibinfo{year}{1966}\natexlab{}.
\newblock \showarticletitle{Binary codes capable of correcting deletions, insertions, and reversals}. In \bibinfo{booktitle}{\emph{Soviet physics doklady}}, Vol.~\bibinfo{volume}{10}. \bibinfo{pages}{707--710}.
\newblock


\bibitem[Li and Wu(2006)]%
        {li2006effective}
\bibfield{author}{\bibinfo{person}{Kanglin Li} {and} \bibinfo{person}{Mengqi Wu}.} \bibinfo{year}{2006}\natexlab{}.
\newblock \bibinfo{booktitle}{\emph{Effective GUI testing automation: Developing an automated GUI testing tool}}.
\newblock \bibinfo{publisher}{John Wiley \& Sons}.
\newblock


\bibitem[Lin et~al\mbox{.}(2019a)]%
        {8952228}
\bibfield{author}{\bibinfo{person}{Jun-Wei Lin}, \bibinfo{person}{Reyhaneh Jabbarvand}, {and} \bibinfo{person}{Sam Malek}.} \bibinfo{year}{2019}\natexlab{a}.
\newblock \showarticletitle{Test Transfer Across Mobile Apps Through Semantic Mapping}. In \bibinfo{booktitle}{\emph{2019 34th IEEE/ACM International Conference on Automated Software Engineering (ASE)}}. \bibinfo{pages}{42--53}.
\newblock
\urldef\tempurl%
\url{https://doi.org/10.1109/ASE.2019.00015}
\showDOI{\tempurl}


\bibitem[Lin et~al\mbox{.}(2019b)]%
        {lin2019craftdroid}
\bibfield{author}{\bibinfo{person}{Jun-Wei Lin}, \bibinfo{person}{Reyhaneh Jabbarvand}, {and} \bibinfo{person}{Sam Malek}.} \bibinfo{year}{2019}\natexlab{b}.
\newblock \showarticletitle{Test Transfer Across Mobile Apps Through Semantic Mapping}. In \bibinfo{booktitle}{\emph{34th International Conference on Automated Software Engineering (ASE 2019)}}.
\newblock


\bibitem[Lin and Malek(2022)]%
        {lin2022gui}
\bibfield{author}{\bibinfo{person}{Jun-Wei Lin} {and} \bibinfo{person}{Sam Malek}.} \bibinfo{year}{2022}\natexlab{}.
\newblock \showarticletitle{Gui test transfer from web to android}. In \bibinfo{booktitle}{\emph{2022 IEEE Conference on Software Testing, Verification and Validation (ICST)}}. IEEE, \bibinfo{pages}{1--11}.
\newblock


\bibitem[Linares-V{\'a}squez et~al\mbox{.}(2017)]%
        {linares2017developers}
\bibfield{author}{\bibinfo{person}{Mario Linares-V{\'a}squez}, \bibinfo{person}{Carlos Bernal-C{\'a}rdenas}, \bibinfo{person}{Kevin Moran}, {and} \bibinfo{person}{Denys Poshyvanyk}.} \bibinfo{year}{2017}\natexlab{}.
\newblock \showarticletitle{How do developers test android applications?}. In \bibinfo{booktitle}{\emph{2017 IEEE International Conference on Software Maintenance and Evolution (ICSME)}}.
\newblock


\bibitem[Linares-V\'{a}squez et~al\mbox{.}(2015)]%
        {MSR15LanguageModelGUI}
\bibfield{author}{\bibinfo{person}{Mario Linares-V\'{a}squez}, \bibinfo{person}{Martin White}, \bibinfo{person}{Carlos Bernal-C\'{a}rdenas}, \bibinfo{person}{Kevin Moran}, {and} \bibinfo{person}{Denys Poshyvanyk}.} \bibinfo{year}{2015}\natexlab{}.
\newblock \showarticletitle{Mining Android App Usages for Generating Actionable GUI-based Execution Scenarios}. In \bibinfo{booktitle}{\emph{Proceedings of the 12th Working Conference on Mining Software Repositories}} (Florence, Italy) \emph{(\bibinfo{series}{MSR '15})}. \bibinfo{publisher}{IEEE Press}, \bibinfo{address}{Piscataway, NJ, USA}, \bibinfo{pages}{111--122}.
\newblock
\showISBNx{978-0-7695-5594-2}
\urldef\tempurl%
\url{http://dl.acm.org/citation.cfm?id=2820518.2820534}
\showURL{%
\tempurl}


\bibitem[Liu et~al\mbox{.}(2024d)]%
        {liu2024your}
\bibfield{author}{\bibinfo{person}{Jiawei Liu}, \bibinfo{person}{Chunqiu~Steven Xia}, \bibinfo{person}{Yuyao Wang}, {and} \bibinfo{person}{Lingming Zhang}.} \bibinfo{year}{2024}\natexlab{d}.
\newblock \showarticletitle{Is your code generated by chatgpt really correct? rigorous evaluation of large language models for code generation}.
\newblock \bibinfo{journal}{\emph{Advances in Neural Information Processing Systems}}  \bibinfo{volume}{36} (\bibinfo{year}{2024}).
\newblock


\bibitem[Liu et~al\mbox{.}(2017)]%
        {ICSE17TextInput}
\bibfield{author}{\bibinfo{person}{Peng Liu} {et~al\mbox{.}}} \bibinfo{year}{2017}\natexlab{}.
\newblock \showarticletitle{Automatic Text Input Generation for Mobile Testing}. In \bibinfo{booktitle}{\emph{Proceedings of the 39th International Conference on Software Engineering}} (Buenos Aires, Argentina) \emph{(\bibinfo{series}{ICSE '17})}. \bibinfo{publisher}{IEEE Press}, \bibinfo{address}{Piscataway, NJ, USA}, \bibinfo{pages}{643--653}.
\newblock
\showISBNx{978-1-5386-3868-2}
\urldef\tempurl%
\url{https://doi.org/10.1109/ICSE.2017.65}
\showDOI{\tempurl}


\bibitem[Liu et~al\mbox{.}(2022)]%
        {liu2022test}
\bibfield{author}{\bibinfo{person}{Shuqi Liu} {et~al\mbox{.}}} \bibinfo{year}{2022}\natexlab{}.
\newblock \showarticletitle{Test reuse based on adaptive semantic matching across android mobile applications}. In \bibinfo{booktitle}{\emph{2022 IEEE 22nd International Conference on Software Quality, Reliability and Security (QRS)}}. IEEE, \bibinfo{pages}{703--709}.
\newblock


\bibitem[Liu et~al\mbox{.}(2024b)]%
        {liu2024enhancing}
\bibfield{author}{\bibinfo{person}{Shuqi Liu} {et~al\mbox{.}}} \bibinfo{year}{2024}\natexlab{b}.
\newblock \showarticletitle{Enhancing test reuse with GUI events deduplication and adaptive semantic matching}.
\newblock \bibinfo{journal}{\emph{Science of Computer Programming}}  \bibinfo{volume}{232} (\bibinfo{year}{2024}), \bibinfo{pages}{103052}.
\newblock


\bibitem[Liu et~al\mbox{.}(2023)]%
        {liu2023fill}
\bibfield{author}{\bibinfo{person}{Zhe Liu} {et~al\mbox{.}}} \bibinfo{year}{2023}\natexlab{}.
\newblock \showarticletitle{Fill in the blank: Context-aware automated text input generation for mobile gui testing. In 2023 IEEE/ACM 45th International Conference on Software Engineering (ICSE)}.
\newblock \bibinfo{journal}{\emph{IEEE, 1355{\'s}1367}} (\bibinfo{year}{2023}).
\newblock


\bibitem[Liu et~al\mbox{.}(2024c)]%
        {liu2024vision}
\bibfield{author}{\bibinfo{person}{Zhe Liu} {et~al\mbox{.}}} \bibinfo{year}{2024}\natexlab{c}.
\newblock \showarticletitle{Vision-driven Automated Mobile GUI Testing via Multimodal Large Language Model}.
\newblock \bibinfo{journal}{\emph{arXiv preprint arXiv:2407.03037}} (\bibinfo{year}{2024}).
\newblock


\bibitem[Liu et~al\mbox{.}(2024a)]%
        {liu2024make}
\bibfield{author}{\bibinfo{person}{Zhe Liu}, \bibinfo{person}{Chunyang Chen}, \bibinfo{person}{Junjie Wang}, \bibinfo{person}{Mengzhuo Chen}, \bibinfo{person}{Boyu Wu}, \bibinfo{person}{Xing Che}, \bibinfo{person}{Dandan Wang}, {and} \bibinfo{person}{Qing Wang}.} \bibinfo{year}{2024}\natexlab{a}.
\newblock \showarticletitle{Make llm a testing expert: Bringing human-like interaction to mobile gui testing via functionality-aware decisions}. In \bibinfo{booktitle}{\emph{Proceedings of the IEEE/ACM 46th International Conference on Software Engineering}}. \bibinfo{pages}{1--13}.
\newblock


\bibitem[Machiry et~al\mbox{.}(2013)]%
        {DynoDroid}
\bibfield{author}{\bibinfo{person}{Aravind Machiry}, \bibinfo{person}{Rohan Tahiliani}, {and} \bibinfo{person}{Mayur Naik}.} \bibinfo{year}{2013}\natexlab{}.
\newblock \showarticletitle{Dynodroid: An Input Generation System for Android Apps}. In \bibinfo{booktitle}{\emph{Proceedings of the 2013 9th Joint Meeting on Foundations of Software Engineering}} (Saint Petersburg, Russia) \emph{(\bibinfo{series}{ESEC/FSE 2013})}. \bibinfo{publisher}{ACM}, \bibinfo{address}{New York, NY, USA}, \bibinfo{pages}{224--234}.
\newblock
\showISBNx{978-1-4503-2237-9}
\urldef\tempurl%
\url{https://doi.org/10.1145/2491411.2491450}
\showDOI{\tempurl}


\bibitem[Mahmood et~al\mbox{.}(2014)]%
        {EvoDroid}
\bibfield{author}{\bibinfo{person}{Riyadh Mahmood}, \bibinfo{person}{Nariman Mirzaei}, {and} \bibinfo{person}{Sam Malek}.} \bibinfo{year}{2014}\natexlab{}.
\newblock \showarticletitle{EvoDroid: Segmented Evolutionary Testing of Android Apps}. In \bibinfo{booktitle}{\emph{Proceedings of the 22Nd ACM SIGSOFT International Symposium on Foundations of Software Engineering}} (Hong Kong, China) \emph{(\bibinfo{series}{FSE 2014})}. \bibinfo{publisher}{ACM}, \bibinfo{address}{New York, NY, USA}, \bibinfo{pages}{599--609}.
\newblock
\showISBNx{978-1-4503-3056-5}
\urldef\tempurl%
\url{https://doi.org/10.1145/2635868.2635896}
\showDOI{\tempurl}


\bibitem[Mao et~al\mbox{.}(2016a)]%
        {mao2016sapienz}
\bibfield{author}{\bibinfo{person}{Ke Mao}, \bibinfo{person}{Mark Harman}, {and} \bibinfo{person}{Yue Jia}.} \bibinfo{year}{2016}\natexlab{a}.
\newblock \showarticletitle{Sapienz: Multi-objective automated testing for android applications}. In \bibinfo{booktitle}{\emph{Proceedings of the 25th International Symposium on Software Testing and Analysis}}. \bibinfo{pages}{94--105}.
\newblock


\bibitem[Mao et~al\mbox{.}(2016b)]%
        {Sapienz}
\bibfield{author}{\bibinfo{person}{Ke Mao}, \bibinfo{person}{Mark Harman}, {and} \bibinfo{person}{Yue Jia}.} \bibinfo{year}{2016}\natexlab{b}.
\newblock \showarticletitle{Sapienz: Multi-objective Automated Testing for Android Applications}. In \bibinfo{booktitle}{\emph{Proceedings of the 25th International Symposium on Software Testing and Analysis}} (Saarbr\&\#252;cken, Germany) \emph{(\bibinfo{series}{ISSTA 2016})}. \bibinfo{publisher}{ACM}, \bibinfo{address}{New York, NY, USA}, \bibinfo{pages}{94--105}.
\newblock
\showISBNx{978-1-4503-4390-9}
\urldef\tempurl%
\url{https://doi.org/10.1145/2931037.2931054}
\showDOI{\tempurl}


\bibitem[Mariani et~al\mbox{.}(2021)]%
        {mariani2021evolutionary}
\bibfield{author}{\bibinfo{person}{Leonardo Mariani}, \bibinfo{person}{Mauro Pezz{\`e}}, \bibinfo{person}{Valerio Terragni}, {and} \bibinfo{person}{Daniele Zuddas}.} \bibinfo{year}{2021}\natexlab{}.
\newblock \showarticletitle{An Evolutionary Approach to Adapt Tests Across Mobile Apps}. In \bibinfo{booktitle}{\emph{The 2nd ACM/IEEE International Conference on Automation of Software Test (AST 2021)}}.
\newblock


\bibitem[Mariani et~al\mbox{.}(2018)]%
        {mariani2018augusto}
\bibfield{author}{\bibinfo{person}{Leonardo Mariani}, \bibinfo{person}{Mauro Pezz{\`e}}, {and} \bibinfo{person}{Daniele Zuddas}.} \bibinfo{year}{2018}\natexlab{}.
\newblock \showarticletitle{Augusto: Exploiting popular functionalities for the generation of semantic gui tests with oracles}. In \bibinfo{booktitle}{\emph{Proceedings of the 40th International Conference on Software Engineering}}. \bibinfo{pages}{280--290}.
\newblock


\bibitem[Mirzaei et~al\mbox{.}(2015)]%
        {SIG-Droid}
\bibfield{author}{\bibinfo{person}{N. Mirzaei}, \bibinfo{person}{H. Bagheri}, \bibinfo{person}{R. Mahmood}, {and} \bibinfo{person}{S. Malek}.} \bibinfo{year}{2015}\natexlab{}.
\newblock \showarticletitle{SIG-Droid: Automated system input generation for Android applications}. In \bibinfo{booktitle}{\emph{2015 IEEE 26th International Symposium on Software Reliability Engineering (ISSRE)}}. \bibinfo{pages}{461--471}.
\newblock
\urldef\tempurl%
\url{https://doi.org/10.1109/ISSRE.2015.7381839}
\showDOI{\tempurl}


\bibitem[Mishra et~al\mbox{.}(2023)]%
        {mishra2023image}
\bibfield{author}{\bibinfo{person}{Yash Mishra} {et~al\mbox{.}}} \bibinfo{year}{2023}\natexlab{}.
\newblock \showarticletitle{Image Understanding of GUI Widgets for Test Reuse}. In \bibinfo{booktitle}{\emph{2023 3rd International Conference on Pervasive Computing and Social Networking (ICPCSN)}}. IEEE, \bibinfo{pages}{572--579}.
\newblock


\bibitem[Moran et~al\mbox{.}(2016)]%
        {Moran2016CrashScope}
\bibfield{author}{\bibinfo{person}{Kevin Moran} {et~al\mbox{.}}} \bibinfo{year}{2016}\natexlab{}.
\newblock \showarticletitle{Automatically Discovering, Reporting and Reproducing Android Application Crashes}. In \bibinfo{booktitle}{\emph{2016 IEEE International Conference on Software Testing, Verification and Validation (ICST)}}. \bibinfo{pages}{33--44}.
\newblock
\urldef\tempurl%
\url{https://doi.org/10.1109/ICST.2016.34}
\showDOI{\tempurl}


\bibitem[OpenAI({[n.\,d.]})]%
        {GPT-4o}
\bibfield{author}{\bibinfo{person}{OpenAI}.} \bibinfo{year}{[n.\,d.]}\natexlab{}.
\newblock \bibinfo{title}{GPT-4o}.
\newblock \bibinfo{howpublished}{\url{https://platform.openai.com/docs/models/gpt-4o/}}.
\newblock


\bibitem[Qin et~al\mbox{.}(2019)]%
        {ISSTA19TestMig}
\bibfield{author}{\bibinfo{person}{Xue Qin}, \bibinfo{person}{Hao Zhong}, {and} \bibinfo{person}{Xiaoyin Wang}.} \bibinfo{year}{2019}\natexlab{}.
\newblock \showarticletitle{TestMig: Migrating GUI Test Cases from iOS to Android}. In \bibinfo{booktitle}{\emph{Proceedings of the 28th ACM SIGSOFT International Symposium on Software Testing and Analysis}} (Beijing, China) \emph{(\bibinfo{series}{ISSTA 2019})}. \bibinfo{publisher}{ACM}, \bibinfo{address}{New York, NY, USA}, \bibinfo{pages}{284--295}.
\newblock
\showISBNx{978-1-4503-6224-5}
\urldef\tempurl%
\url{https://doi.org/10.1145/3293882.3330575}
\showDOI{\tempurl}


\bibitem[Rau et~al\mbox{.}(2018)]%
        {rau2018transferring}
\bibfield{author}{\bibinfo{person}{Andreas Rau}, \bibinfo{person}{Jenny Hotzkow}, {and} \bibinfo{person}{Andreas Zeller}.} \bibinfo{year}{2018}\natexlab{}.
\newblock \showarticletitle{Transferring tests across web applications}. In \bibinfo{booktitle}{\emph{International Conference on Web Engineering}}. Springer, \bibinfo{pages}{50--64}.
\newblock


\bibitem[Su et~al\mbox{.}(2017a)]%
        {Stoat}
\bibfield{author}{\bibinfo{person}{Ting Su} {et~al\mbox{.}}} \bibinfo{year}{2017}\natexlab{a}.
\newblock \showarticletitle{Guided, Stochastic Model-based GUI Testing of Android Apps}. In \bibinfo{booktitle}{\emph{Proceedings of the 2017 11th Joint Meeting on Foundations of Software Engineering}} (Paderborn, Germany) \emph{(\bibinfo{series}{ESEC/FSE 2017})}. \bibinfo{publisher}{ACM}, \bibinfo{address}{New York, NY, USA}, \bibinfo{pages}{245--256}.
\newblock
\showISBNx{978-1-4503-5105-8}
\urldef\tempurl%
\url{https://doi.org/10.1145/3106237.3106298}
\showDOI{\tempurl}


\bibitem[Su et~al\mbox{.}(2017b)]%
        {su2017guided}
\bibfield{author}{\bibinfo{person}{Ting Su}, \bibinfo{person}{Guozhu Meng}, \bibinfo{person}{Yuting Chen}, \bibinfo{person}{Ke Wu}, \bibinfo{person}{Weiming Yang}, \bibinfo{person}{Yao Yao}, \bibinfo{person}{Geguang Pu}, \bibinfo{person}{Yang Liu}, {and} \bibinfo{person}{Zhendong Su}.} \bibinfo{year}{2017}\natexlab{b}.
\newblock \showarticletitle{Guided, stochastic model-based GUI testing of Android apps}. In \bibinfo{booktitle}{\emph{Proceedings of the 2017 11th Joint Meeting on Foundations of Software Engineering}}. \bibinfo{pages}{245--256}.
\newblock


\bibitem[Sun et~al\mbox{.}(2024)]%
        {sun2024adaplanner}
\bibfield{author}{\bibinfo{person}{Haotian Sun} {et~al\mbox{.}}} \bibinfo{year}{2024}\natexlab{}.
\newblock \showarticletitle{Adaplanner: Adaptive planning from feedback with language models}.
\newblock \bibinfo{journal}{\emph{Advances in Neural Information Processing Systems}}  \bibinfo{volume}{36} (\bibinfo{year}{2024}).
\newblock


\bibitem[Talebipour et~al\mbox{.}(2022)]%
        {mapit}
\bibfield{author}{\bibinfo{person}{Saghar Talebipour}, \bibinfo{person}{Yixue Zhao}, \bibinfo{person}{Luka Dojcilovi\'{c}}, \bibinfo{person}{Chenggang Li}, {and} \bibinfo{person}{Nenad Medvidovi\'{c}}.} \bibinfo{year}{2022}\natexlab{}.
\newblock \showarticletitle{UI test migration across mobile platforms}. In \bibinfo{booktitle}{\emph{Proceedings of the 36th IEEE/ACM International Conference on Automated Software Engineering}} (Melbourne, Australia) \emph{(\bibinfo{series}{ASE '21})}. \bibinfo{publisher}{IEEE Press}, \bibinfo{pages}{756–767}.
\newblock
\showISBNx{9781665403375}
\urldef\tempurl%
\url{https://doi.org/10.1109/ASE51524.2021.9678643}
\showDOI{\tempurl}


\bibitem[Team(2024)]%
        {uiautomator}
\bibfield{author}{\bibinfo{person}{Appium Team}.} \bibinfo{year}{2024}\natexlab{}.
\newblock \bibinfo{title}{Appium UiAutomator2 Driver}.
\newblock \bibinfo{howpublished}{\url{https://github.com/appium/appium-uiautomator2-driver}}.
\newblock


\bibitem[Team(2023)]%
        {Monkey}
\bibfield{author}{\bibinfo{person}{Android~Studio Team}.} \bibinfo{year}{2023}\natexlab{}.
\newblock \bibinfo{title}{UI/Application Exerciser Monkey}.
\newblock \bibinfo{howpublished}{\url{https://developer.android.com/studio/test/monkey}}.
\newblock


\bibitem[Wang et~al\mbox{.}(2024b)]%
        {wang2024feedback}
\bibfield{author}{\bibinfo{person}{Dingbang Wang}, \bibinfo{person}{Yu Zhao}, \bibinfo{person}{Sidong Feng}, \bibinfo{person}{Zhaoxu Zhang}, \bibinfo{person}{William~GJ Halfond}, \bibinfo{person}{Chunyang Chen}, \bibinfo{person}{Xiaoxia Sun}, \bibinfo{person}{Jiangfan Shi}, {and} \bibinfo{person}{Tingting Yu}.} \bibinfo{year}{2024}\natexlab{b}.
\newblock \showarticletitle{Feedback-driven automated whole bug report reproduction for android apps}. In \bibinfo{booktitle}{\emph{Proceedings of the 33rd ACM SIGSOFT International Symposium on Software Testing and Analysis}}. \bibinfo{pages}{1048--1060}.
\newblock


\bibitem[Wang et~al\mbox{.}(2020)]%
        {wang2020combodroid}
\bibfield{author}{\bibinfo{person}{Jue Wang} {et~al\mbox{.}}} \bibinfo{year}{2020}\natexlab{}.
\newblock \showarticletitle{ComboDroid: generating high-quality test inputs for Android apps via use case combinations}. In \bibinfo{booktitle}{\emph{Proceedings of the ACM/IEEE 42nd International Conference on Software Engineering}}. \bibinfo{pages}{469--480}.
\newblock


\bibitem[Wang et~al\mbox{.}(2024a)]%
        {wang2024software}
\bibfield{author}{\bibinfo{person}{Junjie Wang} {et~al\mbox{.}}} \bibinfo{year}{2024}\natexlab{a}.
\newblock \showarticletitle{Software testing with large language models: Survey, landscape, and vision}.
\newblock \bibinfo{journal}{\emph{IEEE Transactions on Software Engineering}} (\bibinfo{year}{2024}).
\newblock


\bibitem[Wei et~al\mbox{.}(2022)]%
        {wei2022chain}
\bibfield{author}{\bibinfo{person}{Jason Wei}, \bibinfo{person}{Xuezhi Wang}, \bibinfo{person}{Dale Schuurmans}, \bibinfo{person}{Maarten Bosma}, \bibinfo{person}{Fei Xia}, \bibinfo{person}{Ed Chi}, \bibinfo{person}{Quoc~V Le}, \bibinfo{person}{Denny Zhou}, {et~al\mbox{.}}} \bibinfo{year}{2022}\natexlab{}.
\newblock \showarticletitle{Chain-of-thought prompting elicits reasoning in large language models}.
\newblock \bibinfo{journal}{\emph{Advances in neural information processing systems}}  \bibinfo{volume}{35} (\bibinfo{year}{2022}), \bibinfo{pages}{24824--24837}.
\newblock


\bibitem[Wen et~al\mbox{.}(2023)]%
        {wen2023droidbot}
\bibfield{author}{\bibinfo{person}{Hao Wen}, \bibinfo{person}{Hongming Wang}, \bibinfo{person}{Jiaxuan Liu}, {and} \bibinfo{person}{Yuanchun Li}.} \bibinfo{year}{2023}\natexlab{}.
\newblock \showarticletitle{Droidbot-gpt: Gpt-powered ui automation for android}.
\newblock \bibinfo{journal}{\emph{arXiv preprint arXiv:2304.07061}} (\bibinfo{year}{2023}).
\newblock


\bibitem[Yang et~al\mbox{.}(2013)]%
        {FASE13Greybox}
\bibfield{author}{\bibinfo{person}{Wei Yang}, \bibinfo{person}{Mukul~R. Prasad}, {and} \bibinfo{person}{Tao Xie}.} \bibinfo{year}{2013}\natexlab{}.
\newblock \showarticletitle{A Grey-Box Approach for Automated GUI-Model Generation of Mobile Applications}. In \bibinfo{booktitle}{\emph{Fundamental Approaches to Software Engineering}}, \bibfield{editor}{\bibinfo{person}{Vittorio Cortellessa} {and} \bibinfo{person}{D{\'a}niel Varr{\'o}}} (Eds.). \bibinfo{publisher}{Springer Berlin Heidelberg}, \bibinfo{address}{Berlin, Heidelberg}, \bibinfo{pages}{250--265}.
\newblock
\showISBNx{978-3-642-37057-1}


\bibitem[Yoon et~al\mbox{.}(2023)]%
        {yoon2023autonomous}
\bibfield{author}{\bibinfo{person}{Juyeon Yoon}, \bibinfo{person}{Robert Feldt}, {and} \bibinfo{person}{Shin Yoo}.} \bibinfo{year}{2023}\natexlab{}.
\newblock \showarticletitle{Autonomous Large Language Model Agents Enabling Intent-Driven Mobile GUI Testing}.
\newblock \bibinfo{journal}{\emph{arXiv preprint arXiv:2311.08649}} (\bibinfo{year}{2023}).
\newblock


\bibitem[Yu et~al\mbox{.}(2023)]%
        {yu2023llm}
\bibfield{author}{\bibinfo{person}{Shengcheng Yu} {et~al\mbox{.}}} \bibinfo{year}{2023}\natexlab{}.
\newblock \showarticletitle{Llm for test script generation and migration: Challenges, capabilities, and opportunities}. In \bibinfo{booktitle}{\emph{2023 IEEE 23rd International Conference on Software Quality, Reliability, and Security (QRS)}}. IEEE, \bibinfo{pages}{206--217}.
\newblock


\bibitem[Zaeem et~al\mbox{.}(2014)]%
        {ICST14OracleGeneration}
\bibfield{author}{\bibinfo{person}{Razieh~Nokhbeh Zaeem}, \bibinfo{person}{Mukul~R. Prasad}, {and} \bibinfo{person}{Sarfraz Khurshid}.} \bibinfo{year}{2014}\natexlab{}.
\newblock \showarticletitle{Automated Generation of Oracles for Testing User-Interaction Features of Mobile Apps}. In \bibinfo{booktitle}{\emph{Proceedings of the 2014 IEEE International Conference on Software Testing, Verification, and Validation}} \emph{(\bibinfo{series}{ICST '14})}. \bibinfo{publisher}{IEEE Computer Society}, \bibinfo{address}{Washington, DC, USA}, \bibinfo{pages}{183--192}.
\newblock
\showISBNx{978-1-4799-2255-0}
\urldef\tempurl%
\url{https://doi.org/10.1109/ICST.2014.31}
\showDOI{\tempurl}


\bibitem[Zhang and Rountev(2017)]%
        {ICSE17AndroidWear}
\bibfield{author}{\bibinfo{person}{Hailong Zhang} {and} \bibinfo{person}{Atanas Rountev}.} \bibinfo{year}{2017}\natexlab{}.
\newblock \showarticletitle{Analysis and Testing of Notifications in Android Wear Applications}. In \bibinfo{booktitle}{\emph{Proceedings of the 39th International Conference on Software Engineering}} (Buenos Aires, Argentina) \emph{(\bibinfo{series}{ICSE '17})}. \bibinfo{publisher}{IEEE Press}, \bibinfo{address}{Piscataway, NJ, USA}, \bibinfo{pages}{347--357}.
\newblock
\showISBNx{978-1-5386-3868-2}
\urldef\tempurl%
\url{https://doi.org/10.1109/ICSE.2017.39}
\showDOI{\tempurl}


\bibitem[Zhang et~al\mbox{.}(2024a)]%
        {temdroid}
\bibfield{author}{\bibinfo{person}{Yakun Zhang} {et~al\mbox{.}}} \bibinfo{year}{2024}\natexlab{a}.
\newblock \showarticletitle{Learning-based Widget Matching for Migrating GUI Test Cases}. In \bibinfo{booktitle}{\emph{Proceedings of the IEEE/ACM 46th International Conference on Software Engineering}} (Lisbon, Portugal) \emph{(\bibinfo{series}{ICSE '24})}. \bibinfo{publisher}{Association for Computing Machinery}, \bibinfo{address}{New York, NY, USA}, Article \bibinfo{articleno}{69}, \bibinfo{numpages}{13}~pages.
\newblock
\showISBNx{9798400702174}
\urldef\tempurl%
\url{https://doi.org/10.1145/3597503.3623322}
\showDOI{\tempurl}


\bibitem[Zhang et~al\mbox{.}(2024b)]%
        {zhang2024llmbasedabstractionconcretizationgui}
\bibfield{author}{\bibinfo{person}{Yakun Zhang} {et~al\mbox{.}}} \bibinfo{year}{2024}\natexlab{b}.
\newblock \bibinfo{title}{LLM-based Abstraction and Concretization for GUI Test Migration}.
\newblock
\newblock
\showeprint[arxiv]{2409.05028}~[cs.SE]
\urldef\tempurl%
\url{https://arxiv.org/abs/2409.05028}
\showURL{%
\tempurl}


\bibitem[Zhang et~al\mbox{.}(2024c)]%
        {zhang2024synthesis}
\bibfield{author}{\bibinfo{person}{Yakun Zhang} {et~al\mbox{.}}} \bibinfo{year}{2024}\natexlab{c}.
\newblock \showarticletitle{Synthesis-Based Enhancement for GUI Test Case Migration}. In \bibinfo{booktitle}{\emph{Proceedings of the 33rd ACM SIGSOFT International Symposium on Software Testing and Analysis}}. \bibinfo{pages}{869--881}.
\newblock


\bibitem[Zhao et~al\mbox{.}(2019a)]%
        {zhao2019recdroid}
\bibfield{author}{\bibinfo{person}{Yu Zhao} {et~al\mbox{.}}} \bibinfo{year}{2019}\natexlab{a}.
\newblock \showarticletitle{Recdroid: automatically reproducing android application crashes from bug reports}. In \bibinfo{booktitle}{\emph{2019 IEEE/ACM 41st International Conference on Software Engineering (ICSE)}}. IEEE, \bibinfo{pages}{128--139}.
\newblock


\bibitem[Zhao et~al\mbox{.}(2020)]%
        {Zhao_2020}
\bibfield{author}{\bibinfo{person}{Yixue Zhao}, \bibinfo{person}{Justin Chen}, \bibinfo{person}{Adriana Sejfia}, \bibinfo{person}{Marcelo Schmitt~Laser}, \bibinfo{person}{Jie Zhang}, \bibinfo{person}{Federica Sarro}, \bibinfo{person}{Mark Harman}, {and} \bibinfo{person}{Nenad Medvidovic}.} \bibinfo{year}{2020}\natexlab{}.
\newblock \showarticletitle{FrUITeR: a framework for evaluating UI test reuse}. In \bibinfo{booktitle}{\emph{Proceedings of the 28th ACM Joint Meeting on European Software Engineering Conference and Symposium on the Foundations of Software Engineering}} \emph{(\bibinfo{series}{ESEC/FSE ’20})}. \bibinfo{publisher}{ACM}.
\newblock
\urldef\tempurl%
\url{https://doi.org/10.1145/3368089.3409708}
\showDOI{\tempurl}


\bibitem[Zhao et~al\mbox{.}(2019b)]%
        {zhao2019automatically}
\bibfield{author}{\bibinfo{person}{Yu Zhao}, \bibinfo{person}{Kye Miller}, \bibinfo{person}{Tingting Yu}, \bibinfo{person}{Wei Zheng}, {and} \bibinfo{person}{Minchao Pu}.} \bibinfo{year}{2019}\natexlab{b}.
\newblock \showarticletitle{Automatically extracting bug reproducing steps from android bug reports}. In \bibinfo{booktitle}{\emph{Reuse in the Big Data Era: 18th International Conference on Software and Systems Reuse, ICSR 2019, Cincinnati, OH, USA, June 26--28, 2019, Proceedings 18}}. Springer, \bibinfo{pages}{100--111}.
\newblock


\bibitem[Zheng et~al\mbox{.}(2024)]%
        {zheng2024gpt}
\bibfield{author}{\bibinfo{person}{Boyuan Zheng} {et~al\mbox{.}}} \bibinfo{year}{2024}\natexlab{}.
\newblock \showarticletitle{Gpt-4v (ision) is a generalist web agent, if grounded}.
\newblock \bibinfo{journal}{\emph{arXiv preprint arXiv:2401.01614}} (\bibinfo{year}{2024}).
\newblock


\end{thebibliography}
\end{document}